\documentclass[12pt,a4paper]{article}

\usepackage{amsmath,amssymb,amsthm}
\usepackage{mathtools}

\usepackage[margin=2.5cm]{geometry}
\usepackage{setspace}
\usepackage{parskip}

\usepackage{graphicx}
\graphicspath{{./}}  %

\usepackage{booktabs}
\usepackage{array}

\usepackage{enumitem}

\usepackage[colorlinks=true,linkcolor=blue,citecolor=blue,urlcolor=blue]{hyperref}
\usepackage[capitalise,noabbrev]{cleveref}

\usepackage[authoryear,round]{natbib}
\newtheorem{theorem}{Theorem}
\newtheorem{proposition}{Proposition}

\theoremstyle{definition}

\theoremstyle{remark}
\newtheorem*{remark}{Remark}

\newcommand{\tw}{\tau_w}
\newcommand{\dd}{\mathrm{d}}

\newcommand{\EE}{\mathbb{E}}
\newcommand{\Gini}{\mathrm{Gini}}

\title{Tax Migration as Social Contagion:\\
A Tipping-Point Model with Application to the Scandinavian Wealth Tax Debate}
\author{Anders G.\ Fr{\o}seth}
\date{\today}

\begin{document}
\maketitle

\begin{abstract}
\noindent
\citet{Blandhol2025} estimates that wealth-tax-induced emigration from
Norway reduces long-run GDP by 1.3\%.  Dansk Industri scaled this
figure to argue that a Danish wealth tax would cost billions---a claim
central to the 2026 Danish election campaign.  We develop a social
contagion model in which the emigration rate depends on a
visibility-weighted fraction of prior emigrants, producing
tipping-point dynamics.  Embedding the model in the Fokker--Planck
framework for wealth distributions, we show that the micro-to-macro
extrapolation underlying the 1.3\% figure requires five identification
conditions to hold simultaneously---each of which is violated.  Using a
panel of the 400 wealthiest Norwegians (2011--2025), we estimate the
Pareto tail exponent ($\hat{\alpha} \approx 1.3$, stable across years),
identify the emigrants within Blandhol's sample window (2016--2020),
and document a hidden channel of heir-emigration---36 recent cases
carrying approximately 127~bn~NOK---invisible in the panel data because
controlling owners retain their A-shares while heirs emigrate with
economic exposure only.  The event-study sample is dominated by passive
wealth-holders with near-zero productivity haircuts, and the
wealth-weighted integral that determines the GDP effect is controlled by
individuals entirely absent from the sample.  The Norwegian emigration
wave is a non-scalable, path-dependent tipping event, not a smooth
elasticity.


\end{abstract}

\subsection*{Notation}

\noindent
The following table summarises the principal symbols.  Symbols marked
\dag\ are specific to the contagion model developed in this paper;
the remainder follow the conventions of the companion papers
\citep{Froeseth2026N, Froeseth2026S, Froeseth2026R}.

\medskip
\begin{center}
\footnotesize
\renewcommand{\arraystretch}{1.25}
\begin{tabular}{@{}lp{9.5cm}l@{}}
\toprule
\textbf{Symbol} & \textbf{Description} & \textbf{Eq.} \\
\midrule
\multicolumn{3}{@{}l}{\textit{Contagion model}} \\[2pt]
$n(t)$ & Emigrant fraction of the ultra-wealthy population\dag & \eqref{eq:main} \\
$\tilde{n}(t)$ & Visibility-weighted emigrant fraction\dag & \eqref{eq:ntilde} \\
$\lambda(\tilde{n},t)$ & Effective emigration rate\dag & \eqref{eq:lambda} \\
$\lambda_0(t)$ & Baseline emigration rate (tax $+$ regime)\dag & \eqref{eq:lambda0} \\
$\rho$ & Return migration rate\dag & \eqref{eq:main} \\
$\kappa$ & Contagion strength\dag & \eqref{eq:phi} \\
$\Phi(\tilde{n})$ & Social contagion multiplier,
  $\Phi = e^{\kappa\tilde{n}}$\dag & \eqref{eq:phi} \\
$\xi$ & Visibility (celebrity-effect) exponent\dag & \eqref{eq:ntilde} \\
$\ell$ & Total non-contagion push rate,
  $\ell = \lambda_0(1+\vartheta\,p)$\dag & \eqref{eq:ss_exp} \\
$\vartheta$ & Exit-tax anticipation parameter\dag & \eqref{eq:lambda} \\
$\gamma$ & Sensitivity to reference-dependent tax changes\dag & \eqref{eq:lambda0} \\
$\delta_h$ & Sensitivity to policy-regime hostility\dag & \eqref{eq:lambda0} \\
$h(t)$ & Policy hostility index\dag & \eqref{eq:lambda0} \\
$p(t)$ & Subjective exit-tax probability\dag & \eqref{eq:lambda} \\
$\eta_i$ & Productivity haircut of emigrant~$i$\dag & \eqref{eq:output_general} \\
$\omega_i$ & Capital share of emigrant~$i$\dag & \eqref{eq:output_general} \\
\midrule
\multicolumn{3}{@{}l}{\textit{Fokker--Planck framework (companion papers)}} \\[2pt]
$W$, $x = \ln W$ & Wealth and log-wealth & \eqref{eq:gbm} \\
$\mu$, $\sigma$ & Drift and volatility of wealth returns & \eqref{eq:gbm} \\
$v = \mu - \sigma^2/2$ & Drift in log-wealth & \eqref{eq:logwealth} \\
$D = \sigma^2/2$ & Diffusion coefficient & \eqref{eq:fp} \\
$\tw$ & Wealth tax rate & \eqref{eq:lambda0} \\
$\delta$ & Demographic turnover rate & \eqref{eq:fp_turnover} \\
$\phi(x)$ & Entrant (newborn) distribution & \eqref{eq:fp_turnover} \\
$\zeta$ ($\alpha$) & Pareto tail exponent & \eqref{eq:pareto} \\
$\Lambda$, $t_{1/2}$ & Spectral gap; relaxation half-life & \eqref{eq:halflife} \\
$\theta$ & Book-to-market ratio & --- \\
$\beta$ & Statutory assessment fraction & \eqref{eq:ch1} \\
$\alpha_K$ & Capital share in production & \eqref{eq:output_general} \\
\bottomrule
\end{tabular}
\end{center}

\newpage

\section{Introduction}\label{sec:intro}

In March 2026, wealth taxation became one of the most contentious
issues of the Danish general election.  The Social Democrats proposed a 0.5\% annual tax on
net wealth above 25M DKK for single taxpayers (50M DKK for couples),
estimated to affect approximately 22\,000 taxpayers and raise 6--7
billion DKK per year.  Enhedslisten and SF put forward broader
variants; Fagbev{\ae}gelsens Hovedorganisation (FH) endorsed the
principle.  Opponents, led by Dansk Industri (DI), argued that such a
tax would trigger capital flight and reduce Danish GDP by billions.

The empirical foundation for the opposition's case is a single study.
\citet{Blandhol2025}, in a Princeton job-market paper, exploits the
2022 increase in the Norwegian wealth tax rate to estimate a dynamic
event-study specification showing a 12.6\% average revenue decline in
firms whose owners emigrate.  Through a structural model, she
extrapolates this to a long-run GDP loss of 1.3\%.  The DI analysis
\citep{DI2026} scales the Norwegian estimate to Danish conditions by
comparing the ratio of proposed Danish wealth tax revenue to GDP with
the corresponding Norwegian ratio, yielding projected annual GDP losses
of 11--28 billion DKK depending on the proposal---multiples of the
expected revenue.

This linear cross-country scaling rests on five identification
conditions (\Cref{sec:gap_summary}), each of which we show to be
violated.

We develop a model of tax-motivated emigration as a social contagion
process.  The emigration rate depends not only on the tax differential
but on a visibility-weighted fraction of prior emigrants (the
``celebrity effect''), on reference-dependent felt tax pressure, on
policy regime hostility, and on exit-tax anticipation.  The resulting
scalar ODE admits a saddle-node bifurcation: when the contagion
strength exceeds a critical threshold ($\kappa \geq 4$), two stable
equilibria coexist for an interval of push rates and the system can tip
discontinuously from one to the other.  The Norwegian 2022 episode---in which the
emigration rate jumped from 0.2\% to over 2\%---is consistent with such
a tipping event, not with a smooth elasticity.

Our focus on the migration channel is deliberate.  \citet{Brulhart2022}
decompose the Swiss wealth tax response into three components: taxpayer
migration ($\sim$24\%), capitalisation into housing prices ($\sim$21\%),
and evasion or avoidance ($\sim$55\%).  The present paper isolates the
first channel, because it is the component most sensitive to contagion
dynamics and the one through which the Blandhol GDP-loss argument
operates.  We take no stance on the total wealth-tax elasticity; the
object of interest is the contagion-driven migration component alone.

To assess the empirical validity of the GDP scaling, we construct a
panel of the 400 wealthiest Norwegians over 2011--2025 from the annual
Kapital~400 list.  The data allow us to estimate the Pareto tail
exponent directly ($\hat{\alpha} \approx 1.3$, remarkably stable across
years), to identify which ultra-wealthy individuals actually emigrated
during Blandhol's 2016--2020 sample window, and to classify them by
the nature of their wealth (active control versus passive holdings).
The findings are stark: only seven Kapital~400 members emigrated during
the window, only one ran a domestic operating business, and none of the
top emigrant fortunes---which dominate the wealth-weighted GDP
integral---are represented in the event-study sample.

The paper embeds these findings in the Fokker--Planck framework for
wealth distributions developed in three companion papers
\citep{Froeseth2026N, Froeseth2026S, Froeseth2026R}.  The key results
from that framework---the drift-shift symmetry of proportional wealth
taxation, the Pareto tail exponent, the spectral gap determining
relaxation timescales, and the classification of migration as a
permeable-boundary leakage channel---are summarised self-containedly in
\Cref{sec:fp} and applied throughout.

The paper proceeds as follows.  \Cref{sec:facts} sets out six stylized
facts that motivate the contagion model.  \Cref{sec:model} develops the
model and \Cref{sec:tipping} derives the tipping-point condition.
\Cref{sec:fp} provides the Fokker--Planck foundation.
\Cref{sec:data} presents the empirical evidence from the Kapital~400
panel.  \Cref{sec:gap} uses the framework and data to expose five
identification failures in the micro-to-macro extrapolation.
\Cref{sec:calibration} calibrates the identifiable parameters and
is explicit about what the data cannot pin down.
\Cref{sec:policy} draws implications for the Danish debate, and
\Cref{sec:conclusion} concludes.

\section{Stylized facts}\label{sec:facts}

The Norwegian wealth tax emigration episode (2022--) and comparable
episodes elsewhere suggest that tax migration among the ultra-wealthy
is not well described by independent, rational cost--benefit
calculations.  The following stylized facts motivate the model.

\begin{table}[tp]
\centering
\footnotesize
\renewcommand{\arraystretch}{1.15}
\begin{tabular}{@{}cp{10.5cm}@{}}
\toprule
\textbf{F} & \textbf{Stylized fact} \\
\midrule
F1 & \textbf{Social contagion / ``Keeping up with the Joneses.''}
  The propensity to emigrate depends on what peers have done.
  High-profile first movers lower the social cost and raise the
  perceived status of emigration.  The influence of an individual
  departure scales with the emigrant's visibility (wealth,
  media profile), not just with the head count.
  The mechanism echoes \citet{Abel1990}: utility depends on
  one's position relative to a social benchmark, here the
  emigration decisions of visible peers. \\[2pt]
F2 & \textbf{Reference-dependent tax pressure.}
  The emigration response depends on \emph{changes} in tax levels
  relative to a reference point, not on the absolute level.
  A recent increase triggers a larger response than the same
  level held for decades.  This is the core prediction of prospect
  theory \citep{KahnemanTversky1979}: outcomes are evaluated as
  deviations from a reference point, and losses (tax increases)
  loom larger than equivalent gains. \\[2pt]
F3 & \textbf{Policy regime hostility.}
  The propensity to emigrate responds to the \emph{overall} policy
  environment towards wealthy capital owners---rhetoric, proposed
  reforms, political signals---beyond any single tax parameter.
  The 2021 Norwegian government change was perceived as a regime
  shift. \\[2pt]
F4 & \textbf{Exit tax anticipation and the crystallisation motive.}
  Expectations of stricter exit taxes create a rush-for-the-door
  effect: agents who might have emigrated gradually accelerate
  their departure to avoid being locked in.  The motive is
  amplified when emigrants hold large deferred tax liabilities
  inside holding structures (via participation exemptions such as
  the Norwegian \emph{fritaksmetoden}): emigration before the exit
  tax window closes allows crystallising these gains under a more
  favourable regime.  The Norwegian five-year lapse rule---abolished
  29~November 2022---made this crystallisation opportunity concrete
  and time-limited.  The expectations-based reference point of
  \citet{KoszegiRabin2006} is directly applicable: the anticipated
  future tax regime---not just the current one---shapes the
  perceived loss from staying. \\[2pt]
F5 & \textbf{Wealth--control separation.}
  Most emigrating wealth is \emph{passive}: heirs receive B-shares
  (economic exposure, dividends) while the controlling owner retains A-shares
  (voting rights, strategic control) and professional management
  remains domestic.  Among the 36 recent heir-emigrants identified
  from Kapital~400 and Kapital~300 families
  (\Cref{sec:heir_emigrants}), the combined inherited wealth is
  approximately 127~bn~NOK; in every case the controlling parent or
  grandparent continued to reside in Norway and manage the
  enterprise.  The ``productivity haircut'' is near zero
  for this category. \\[2pt]
F6 & \textbf{Tipping dynamics.}
  Emigration rates can jump discontinuously (0.2\% to $>2\%$ in
  Norway) and cluster geographically in Swiss low-tax locations
  (Lugano, Zug, Schwyz, municipalities in Kanton Z\"urich),
  while higher-tax Geneva is largely avoided.  Some emigrants
  maintain business offices in Stadt Z\"urich---or even reside
  there, trading tax minimisation for urban amenities---but the
  dominant pattern favours low-tax cantons, consistent with a
  phase transition rather than a smooth elasticity. \\
\bottomrule
\end{tabular}
\caption{Stylized facts motivating the contagion model.}\label{tab:stylized_facts}
\end{table}

\medskip
These facts find direct empirical support both within and outside
Norway.  \citet{IaconoSmedsvik2024} exploit the B{\o} municipality
experiment---a single Norwegian municipality unilaterally reduced its
municipal wealth tax component from 0.85\% to 0.35\% in 2021---and find
that taxpayer mobility accounts for roughly 79\% of the resulting
increase in local taxable wealth.  This is a within-Norway proof of
concept for the mobility margin at a modest rate differential.  The
United States contrast is instructive: \citet{Young2016} document that
millionaire interstate migration averages just 2.4\% per year---below
the general population rate of 2.9\%---despite several states levying
no income tax.  The explanation they advance is \emph{embeddedness}:
location-specific social capital, business ties, family attachments,
and lifecycle constraints impose frictions that dominate plausible tax
savings.  The contrast matters for interpretation.  F1--F6 are not
universal; they describe a regime that becomes salient only when
embeddedness frictions are weak, or when a focal shock (F3, F4)
overrides them.  The Nordic ultra-wealthy---geographically mobile,
highly networked, and facing a policy shock bundled with an exit-tax
deadline---appear to fall squarely into that regime.

\section{The model}\label{sec:model}
\subsection{Emigrant fraction dynamics}\label{sec:dynamics}

Let $n(t) \in [0,1]$ denote the fraction of the ultra-wealthy
population that has emigrated.  We model the emigrant fraction as a
scalar ODE:
\begin{equation}\label{eq:main}
  \boxed{\dot{n}
  = \underbrace{\lambda(\tilde{n}, t)}_{\text{push rate}}
    \cdot (1 - n)
    - \underbrace{\rho}_{\text{return rate}}
    \cdot n \,,}
\end{equation}
where $1 - n$ is the resident fraction, $\rho > 0$ is the (constant)
return migration rate, and $\lambda$ is the \emph{effective emigration
rate} specified below.  The factor $(1-n)$ ensures that emigration
depletes the resident pool.

This is a reduced model.  It arises from the full coupled
Fokker--Planck system under a separation-of-scales
assumption: wealth distributions equilibrate fast relative to migration
dynamics, so the migration decision can be studied as a scalar ODE for
the aggregate emigrant fraction (the derivation is given in
\Cref{sec:fp_reduction}).

\subsection{The effective emigration rate}\label{sec:lambda}

The emigration rate has four multiplicative components, each capturing
one stylized fact:
\begin{equation}\label{eq:lambda}
  \lambda(\tilde{n}, t)
  = \underbrace{\lambda_0(t)}_{\substack{\text{baseline:}\\\text{tax + regime}}}
    \cdot\; \underbrace{\Phi(\tilde{n})}_{\substack{\text{social}\\\text{contagion}}}
    \cdot\; \underbrace{(1 + \vartheta\, p(t))}_{\substack{\text{exit-tax}\\\text{anticipation}}} \,.
\end{equation}

\paragraph{Baseline rate (F2 + F3).}
The baseline rate captures both the tax motive and the policy
environment:
\begin{equation}\label{eq:lambda0}
  \lambda_0(t)
  = \bar{\lambda}\;
    \bigl(\tw(t) - \tw^* + \gamma\,\Delta\tw\bigr)
    \cdot (1 + \delta_h\, h(t)) \,,
\end{equation}
where $\tw(t)$ is the statutory wealth tax rate, $\tw^*$ the foreign
rate, $\Delta\tw = [\tw(t) - \tw(t - \Delta)]^+$ the recent tax
increase (reference dependence, F2), and $h(t) \in [0,1]$ the policy
hostility index~(F3).  The $\Delta\tw$ term is the prospect-theoretic
channel: agents evaluate the tax burden as a deviation from a reference
point (the prior rate), and the asymmetric response to increases versus
stable levels reflects loss aversion
\citep{KahnemanTversky1979}.  The parameters $\gamma > 0$ and
$\delta_h > 0$ govern the sensitivity to tax changes and regime
hostility, respectively.  The normalisation constant $\bar{\lambda}$ converts tax
differentials into an emigration rate.

\emph{Note on the effective tax differential.}  The model uses the
wealth tax rate~$\tw$ as the driving variable, but the true tax
differential motivating emigration is a composite of wealth tax,
deferred capital gains tax (accumulated inside holding structures via
participation exemptions), dividend tax, and exit-tax rules.  This
composite enters the model through the calibration
of~$\bar{\lambda}$: a high~$\bar{\lambda}$ reflects a setting where
multiple tax channels reinforce the emigration incentive, as in
Norway.  This matters for cross-country extrapolation: if
$\bar{\lambda}$ is estimated from Norwegian data where the composite
burden is high, applying it to Denmark---where the wealth tax would
be new and no comparable deferral overhang exists---overstates the
baseline emigration rate.

\paragraph{Social contagion (F1).}
The social multiplier amplifies the baseline rate as more agents
emigrate:
\begin{equation}\label{eq:phi}
  \Phi(\tilde{n}) = e^{\kappa\,\tilde{n}} \,,
\end{equation}
where $\kappa > 0$ is the contagion strength.  The exponential form
has three motivations.  First, it is the \emph{hazard-rate}
specification standard in epidemiology
\citep[see][for a survey]{Hethcote2000}: each additional emigrant
adds a constant~$\kappa$ to the log-emigration rate, so that
$\ln\Phi = \kappa\,\tilde{n}$.  Second, it nests the linear
approximation $\Phi \approx 1 + \kappa\,\tilde{n}$ for small
emigrant fractions $\tilde{n} \ll 1/\kappa$ but generates the
nonlinear positive feedback needed for tipping when $\tilde{n}$ is
large (as we show in \Cref{sec:ss}, the linear form cannot produce
bistability).  Third, the exponential form connects to threshold models of
collective behaviour \citep{Granovetter1978,Schelling1971}.
Suppose each individual~$i$ emigrates if and only if
$\tilde{n}$ exceeds a personal threshold~$\theta_i$ drawn from
an exponential distribution with rate~$\kappa$.  The exponential
distribution has a constant hazard rate: among those who have not
yet emigrated, each marginal increase in~$\tilde{n}$ converts a
fixed fraction~$\kappa$ of the remainder.  This constant
proportional conversion rate is precisely the log-linear property
$\dd\ln\Phi/\dd\tilde{n} = \kappa$ that
yields~\eqref{eq:phi}.%
\footnote{Formally, if thresholds have CDF
  $F(\theta) = 1 - e^{-\kappa\theta}$, the fraction that has
  crossed its threshold at level~$\tilde{n}$ is
  $F(\tilde{n}) = 1 - e^{-\kappa\tilde{n}}$---a bounded adoption
  curve.  The social multiplier~$\Phi$ in~\eqref{eq:phi} is a
  different object: it tracks the \emph{rate} at which new
  conversions occur relative to the remaining stock, which is the
  hazard rate $f/(1-F) = \kappa$.  Integrating
  $\dd\ln\Phi = \kappa\,\dd\tilde{n}$ with $\Phi(0) = 1$ gives
  $\Phi(\tilde{n}) = e^{\kappa\tilde{n}}$.  For small
  $\tilde{n}$ both the adoption curve and the multiplier reduce to
  $\approx \kappa\tilde{n}$; they diverge in the nonlinear regime,
  where the multiplier captures the self-reinforcing feedback.}
The specification is also used in the social interactions literature
\citep{Scheinkman2008} and in models of complex contagion where
adoption requires reinforcement from multiple peers
\citep{Centola2018}.

The \emph{visibility-weighted} emigrant fraction is
\begin{equation}\label{eq:ntilde}
  \tilde{n}(t)
  = \frac{\sum_{i \in \text{emigrants}} W_i^{\,\xi}}
         {\sum_{i \in \text{all}} W_i^{\,\xi}} \,,
\end{equation}
where $W_i$ is the wealth of agent~$i$ and $\xi \geq 0$ governs the
celebrity effect (F1).  When $\xi = 0$, $\tilde{n} = n$ (pure head
count).  When $\xi > 0$, a single prominent departure shifts
$\tilde{n}$ far more than many smaller ones.

\paragraph{Exit-tax anticipation (F4).}
The factor $(1 + \vartheta\,p(t))$ accelerates emigration when agents
expect exit taxes.  The parameter $\vartheta > 0$ governs the
anticipatory acceleration and $p(t) \in [0,1]$ is the subjective
exit-tax probability, which may itself depend on $n$ (more emigration
raises political pressure for exit taxes), creating a self-reinforcing
loop.  The reference point here is the \emph{expected} future tax
regime, not just the current one---an expectations-based reference
point in the sense of \citet{KoszegiRabin2006}.

\subsection{Wealth--control separation (F5)}\label{sec:types}

Not all emigrating wealth carries a productivity cost.  We distinguish:
\begin{itemize}[nosep]
  \item \textbf{Active-control owners} ($\mathcal{A}$):
    hold voting rights, exercise strategic oversight.  Productivity
    haircut $\eta_A$ (small when management is delegated to a
    professional CEO).
  \item \textbf{Passive wealth holders} ($\mathcal{P}$):
    hold economic rights (B-shares), no operational role.  Productivity
    haircut $\eta_P \approx 0$ by construction.
\end{itemize}
In the typical Norwegian family firm (dual share classes), the heir
emigrates with the B-shares while the controlling parent retains
A-shares domestically.  The wealth moves; the productive control does
not.

\section{Steady states and tipping}\label{sec:tipping}

The ODE~\eqref{eq:main} is a one-dimensional dynamical system.  For
such systems, the qualitative behaviour---how many steady states
exist and whether the system can jump discontinuously between
them---is determined by the shape of the right-hand side as a
function of the state variable~$n$.  The key phenomenon is
\emph{bistability}: for certain parameter values, two stable
equilibria coexist (low and high emigration), separated by an
unstable one.  A small, slow change in the push rate~$\ell$ can then
cause the system to \emph{tip}---to jump abruptly from the
low-emigration equilibrium to the high-emigration one.  This is the
mechanism behind the ``cascade'' dynamics of
\citet{Granovetter1978}, the ``tipping'' of \citet{Schelling1971},
and the ``revolutionary threshold'' of \citet{Kuran1991}.
Mathematically, the transition is a \emph{saddle-node bifurcation}
(also called a \emph{fold}): a stable and an unstable equilibrium
collide and annihilate each other, leaving only the distant
alternative \citep[ch.~3]{Strogatz2015}.

The clearest recent empirical analogue outside Norway is the Madrid
\emph{efecto Madrid}: when the regional government effectively
abolished its wealth tax while neighbouring Spanish regions retained
theirs, high-wealth migration responded in a near-binary manner rather
than as a smooth function of the tax differential.
\citet{AgrawalEtAl2025} document this pattern across the post-2011
Spanish regional data.  We do not claim their estimates identify
$\kappa$; we note only that the observed discontinuity is
qualitatively consistent with the bistable mechanism derived below.

We now derive the conditions under which this bifurcation occurs.

\subsection{Steady-state equation}\label{sec:ss}

At steady state ($\dot{n} = 0$), \Cref{eq:main} gives
\begin{equation}\label{eq:ss_cond}
  \lambda(\tilde{n}^*)\,(1 - n^*) = \rho\, n^* \,.
\end{equation}
Writing $\ell = \lambda_0\,(1 + \vartheta\,p)$ for the non-contagion
push rate and approximating $\tilde{n} \approx n$ (the $\xi = 0$
case), we substitute $\Phi(n) = e^{\kappa n}$ to obtain the
steady-state condition
\begin{equation}\label{eq:ss_exp}
  e^{\kappa n}\,\ell\,(1 - n) = \rho\, n \,.
\end{equation}
Define the \emph{effective outflow function}
\begin{equation}\label{eq:G}
  G(n) \;\equiv\; e^{\kappa n}\,\ell\,(1 - n) \;-\; \rho\, n \,.
\end{equation}
We have $G(0) = \ell > 0$ and $G(1) = -\rho < 0$, so at least one
interior steady state always exists.  The second derivative is
\begin{equation}\label{eq:Gpp}
  G''(n) = \ell\, e^{\kappa n}\bigl[\kappa^2(1-n) - 2\kappa\bigr],
\end{equation}
which changes sign at $n_{\rm infl} = 1 - 2/\kappa$.  For
$\kappa > 2$, the outflow function is convex near the origin and
concave for $n > n_{\rm infl}$---an S-shape that permits multiple
crossings of the $n$-axis and hence multiple steady states.

\emph{Remark.}  With the linear approximation $\Phi(n) \approx 1 +
\kappa n$, the outflow function becomes a concave quadratic
$G(n) = -\kappa\ell\,n^2 + [(\kappa-1)\ell - \rho]\,n + \ell$,
which has exactly one root in $[0,1]$ for all parameter values.
The linear specification thus produces amplification but not
bistability; the nonlinear (exponential) term is essential for the
fold.

\begin{figure}[tp]
  \centering
  \includegraphics[width=0.55\textwidth]{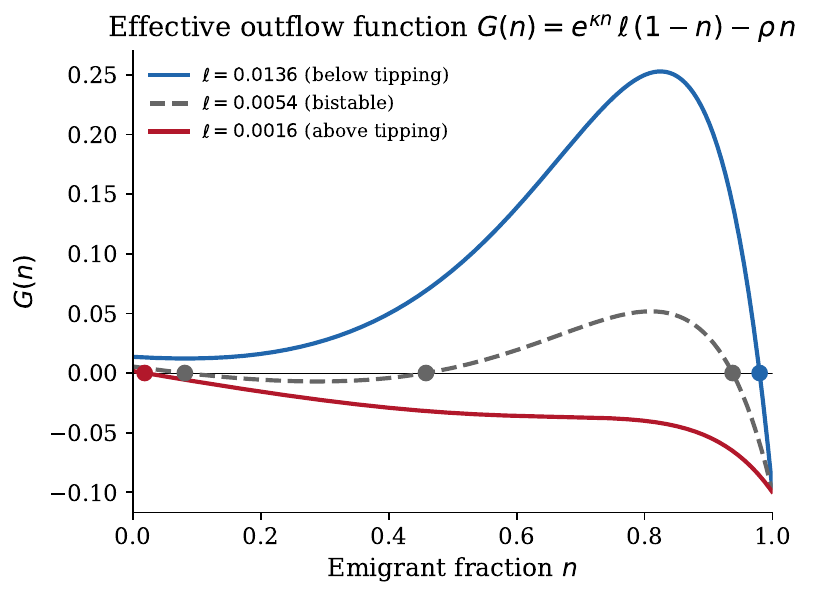}
  \caption{The effective outflow function $G(n) = e^{\kappa n}\,\ell\,(1-n) - \rho\,n$
    for three values of the total push rate~$\ell$.  Zeros of $G$
    (circles) are steady states.  For $\ell$ between the two fold
    values $\ell^-$ and~$\ell^+$, the S-shape created by the
    exponential social multiplier produces three zeros---two stable
    equilibria (low and high emigration) separated by an unstable one.
    Parameters: $\kappa = 6$, $\rho = 0.10$ (schematic).}
  \label{fig:outflow}
\end{figure}

\subsection{Critical contagion strength}\label{sec:critical}

The system transitions from one to three steady states (the
tipping point) when $G$ and $G'$ vanish simultaneously.  Setting
$G'(n) = 0$ gives
\[
  \ell\, e^{\kappa n}\bigl[\kappa(1-n) - 1\bigr] = \rho \,.
\]
Substituting into $G(n) = 0$ and simplifying yields what we call the
\emph{fold condition}---the requirement for a saddle-node bifurcation,
where a stable and an unstable equilibrium collide and
annihilate (see the introductory discussion above)---expressed as a
constraint on $n$:
\begin{equation}\label{eq:fold_condition}
  \kappa\, n^2 - \kappa\, n + 1 = 0 \,.
\end{equation}
This quadratic in~$n$ has real solutions if and only if its
discriminant $\kappa^2 - 4\kappa = \kappa(\kappa - 4)$ is
non-negative, giving:
\begin{equation}\label{eq:kappa_crit}
  \boxed{\kappa^{\mathrm{crit}} = 4 \,.}
\end{equation}

\begin{proposition}[Tipping point]\label{prop:tipping}
For $\kappa < 4$, the system~\eqref{eq:main} has a unique stable
steady state for every~$\ell > 0$.  For $\kappa \geq 4$, there
exists an interval of push rates
$\ell \in (\ell^-(\kappa,\rho),\; \ell^+(\kappa,\rho))$ for which
two stable steady states coexist (low and high emigration),
separated by an unstable equilibrium.  The boundaries $\ell^{\pm}$
are saddle-node bifurcation points.
\end{proposition}%
\footnote{The bifurcation at $\kappa^{\mathrm{crit}} = 4$ is an instance of
a broader class of mean-field threshold phenomena.
\citet{BernardBouchaudLeDoussal2026} find an analogous critical coupling
in the Bouchaud--M\'{e}zard model: below $\varphi_c$, wealth condenses on a
single agent; above it, wealth is delocalised.  The two bifurcations are
mirror images---theirs is a loss of confinement (stabilising coupling too
weak), ours is a loss of cohesion (destabilising coupling too strong)---but
both are controlled by a single dimensionless ratio and exhibit hysteresis.}

At the two saddle-node (fold) bifurcation points, the critical
emigrant fractions are
\begin{equation}\label{eq:n_fold}
  n_{\rm fold}^{\pm}
  = \frac{1 \pm \sqrt{1 - 4/\kappa}}{2} \,,
\end{equation}
and the corresponding critical push rates follow from
\eqref{eq:ss_exp}:
\begin{equation}\label{eq:ell_tip}
  \ell^{\pm}(\kappa,\rho)
  = \frac{\rho\, n_{\rm fold}^{\pm}}
         {e^{\kappa\, n_{\rm fold}^{\pm}}\,
          (1 - n_{\rm fold}^{\pm})} \,.
\end{equation}

\begin{figure}[tp]
  \centering
  \includegraphics[width=0.55\textwidth]{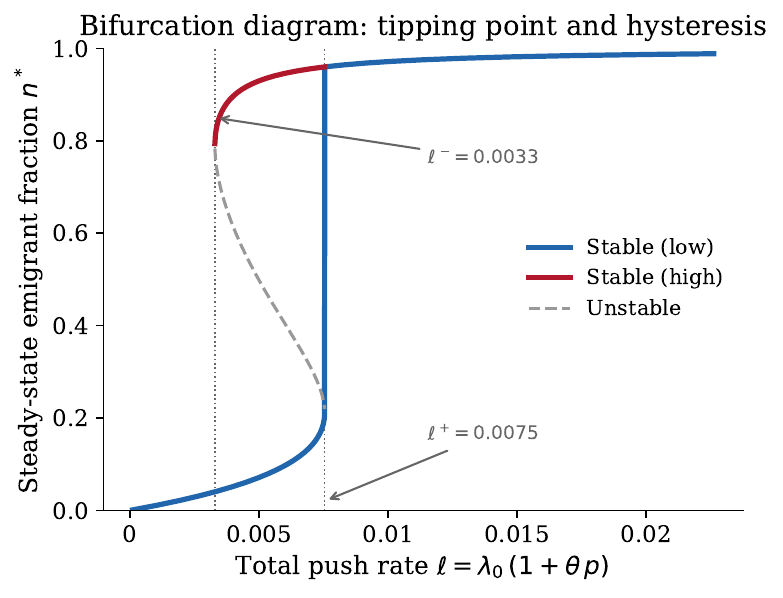}
  \caption{Bifurcation diagram: steady-state emigrant fraction
    $n^*$ as a function of the total push rate~$\ell$.  For
    $\ell \in (\ell^-,\ell^+)$ the system has two stable branches
    (blue: low emigration; red: high emigration) connected by an
    unstable branch (grey dashed).  Increasing $\ell$ past~$\ell^+$
    triggers an irreversible jump to the high-emigration branch;
    recovery requires reducing $\ell$ below~$\ell^- < \ell^+$
    (hysteresis).  Parameters: $\kappa = 6$, $\rho = 0.10$
    (schematic).}
  \label{fig:bifurcation}
\end{figure}

The tipping threshold $\ell^+$ is \emph{decreasing} in~$\kappa$:
stronger contagion means a smaller push suffices to trigger the
cascade.  Conversely, the four destabilising forces raise the
total push rate~$\ell$ toward~$\ell^+$:
\begin{enumerate}[nosep]
  \item A tax increase raises $\ell$ via $\Delta\tw$ (F2).
  \item A hostile policy regime raises $\ell$ via $h$ (F3).
  \item Exit-tax expectations raise $\ell$ via $\vartheta\,p$ (F4).
  \item High visibility ($\xi > 0$) amplifies $\tilde{n}$ relative to
    $n$, effectively increasing the contagion strength (F1).
\end{enumerate}
No single factor needs to be large.  The Norwegian cascade resulted
from the simultaneous activation of all four channels, each
individually insufficient but jointly pushing $\ell$ past the
tipping threshold.

\begin{figure}[tp]
  \centering
  \includegraphics[width=0.55\textwidth]{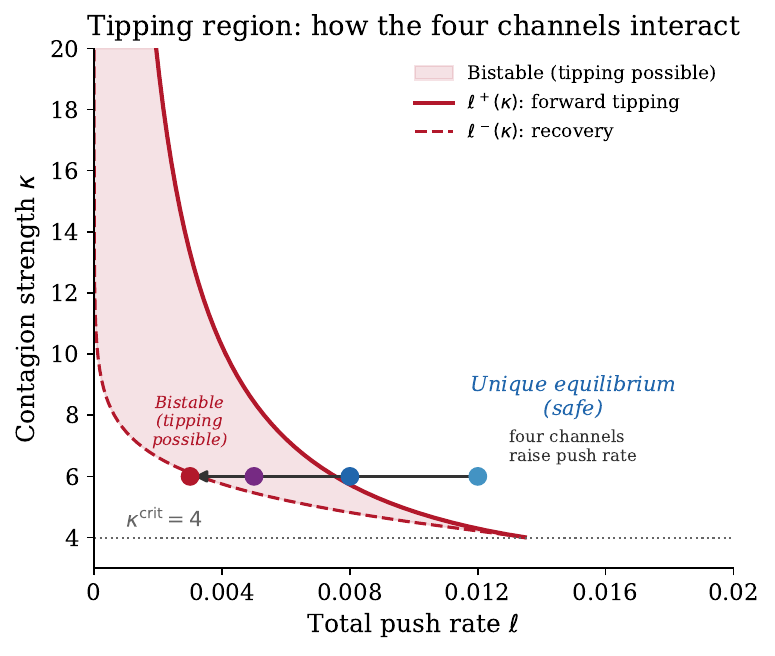}
  \caption{Tipping region in the $(\ell,\kappa)$ plane.  The shaded
    area between the fold curves $\ell^-(\kappa)$ and
    $\ell^+(\kappa)$ is the bistable region where both low- and
    high-emigration steady states coexist.  Below $\kappa^{\rm crit}
    = 4$ (dotted line) no bistability is possible regardless
    of~$\ell$.  The four destabilising channels---tax level,
    reference dependence, policy hostility, and exit-tax
    anticipation---jointly raise the effective push rate toward the
    tipping boundary.  Parameters: $\rho = 0.10$ (schematic).}
  \label{fig:kappa_crit}
\end{figure}

\subsection{The Norwegian narrative}\label{sec:narrative}

Before 2021: $\tw$ stable, $h$ low, $p \approx 0$,
$\tilde{n} \approx 0$.  The total push $\ell$ is small, well below
$\ell^+$---the system sits safely on the low-emigration branch
($n^* \approx 0.2\%$).

2021--2022 sequence:
\begin{enumerate}[nosep]
  \item Change of government raises $h$ sharply (F3).
  \item Wealth tax increase raises $\Delta\tw$; the reference-dependent
    channel amplifies $\ell$ beyond the absolute level (F2).
  \item Discussion of exit taxes raises $p$, activating the
    anticipation multiplier (F4).
  \item A single highly visible departure shifts $\tilde{n}$ far more
    than $n$, triggering the contagion cascade (F1).
\end{enumerate}
The combined effect pushes $\ell$ past $\ell^+$, and the system tips
to the high-emigration branch.  Recovery requires pushing $\ell$
back below $\ell^-\!<\ell^+$ (hysteresis).

Removing any one of the four channels might have been sufficient to
prevent the tipping---which explains why similar tax \emph{levels} in
other Nordic countries have not produced comparable emigration waves.

\begin{figure}[tp]
  \centering
  \includegraphics[width=0.55\textwidth]{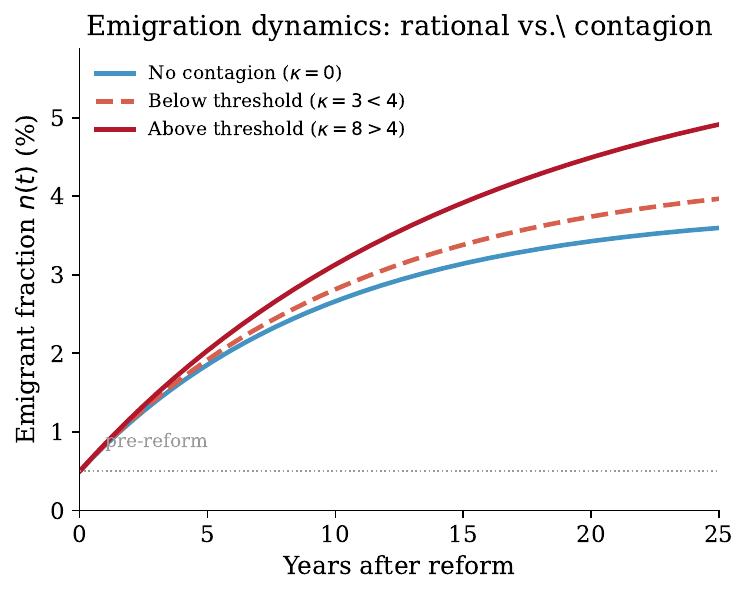}
  \caption{Time trajectories of the emigrant fraction $n(t)$ after a
    permanent increase in the push rate at $t = 0$.  Without
    contagion ($\kappa = 0$) the system converges smoothly to a new
    low equilibrium.  Below the critical threshold ($\kappa = 3 < 4$)
    contagion amplifies the response but the dynamics remain
    monotone.  Above the threshold ($\kappa = 8 > 4$) the system
    undergoes an S-shaped cascade to the high-emigration branch.
    Parameters: $\rho = 0.10$, $\ell = 0.004$ (schematic).}
  \label{fig:trajectories}
\end{figure}

\section{The Fokker--Planck foundation}\label{sec:fp}

This section summarises the results from the Fokker--Planck framework
for wealth distributions developed in \citet{Froeseth2026N,
Froeseth2026S, Froeseth2026R} that are needed for the empirical
analysis.  The presentation is self-contained.

\subsection{Wealth dynamics and the Fokker--Planck equation}\label{sec:fp_gbm}

Individual wealth $W_i(t)$ evolves as geometric Brownian motion:
\begin{equation}\label{eq:gbm}
  \frac{\dd W}{W} = \mu\,\dd t + \sigma\,\dd B_t \,,
\end{equation}
where $\mu$ is the expected return on capital and $\sigma$ the
volatility.  Applying It\^{o}'s lemma to $x = \ln W$ gives:
\begin{equation}\label{eq:logwealth}
  \dd x = v\,\dd t + \sigma\,\dd B_t \,,
  \qquad v \equiv \mu - \tfrac{\sigma^2}{2} \,.
\end{equation}
The probability density $\pi(x,t)$ of log-wealth across the population
satisfies the Fokker--Planck equation:
\begin{equation}\label{eq:fp}
  \frac{\partial \pi}{\partial t}
  = -v\,\frac{\partial \pi}{\partial x}
    + D\,\frac{\partial^2 \pi}{\partial x^2} \,,
  \qquad D = \frac{\sigma^2}{2} \,.
\end{equation}

\subsection{Drift-shift symmetry}\label{sec:drift_shift}

A proportional wealth tax at rate $\tw$ reduces the after-tax return
from $\mu$ to $\mu - \tw$.  In the Fokker--Planck equation, this is a
uniform shift of the drift coefficient:
\begin{equation}\label{eq:drift_shift}
  v \;\to\; v_\tau = v - \tw \,, \qquad D \;\to\; D \,.
\end{equation}
This \emph{drift-shift transformation}
$\mathcal{T}_\tau: v \mapsto v - \tw$, $D \mapsto D$ preserves
drift differences between assets ($v_i - v_j$), diffusion ratios
($D_i/D_j$), and all Sharpe-ratio-like quantities---and therefore
leaves optimal portfolio weights unchanged
\citep[Proposition~2]{Froeseth2026S}.  This is the mathematical
content of wealth tax neutrality: the tax acts as a Galilean-type
boost in log-wealth space.

The neutrality breaks down when the tax base departs from market value.
If asset~$i$ has a book-to-market ratio $\beta_i$, the effective tax
rate is $\tw^{(i)} = \tw \cdot \beta_i$ and the drift shift becomes
asset-dependent:
\begin{equation}\label{eq:ch1}
  v_i \;\to\; v_i - \tw\,\beta_i \,.
\end{equation}
This is the first of several symmetry-breaking channels classified in
\citet{Froeseth2026S}.  In the Norwegian system, unlisted shares are
assessed at book value with a statutory assessment fraction
$\beta = 0.80$, while listed shares are assessed near market value.
The effective tax burden is therefore asset-class-dependent, creating
anisotropic drift across portfolio components.

\begin{theorem}[One-period book-value pricing; {\citet{Froeseth2026N},
Theorem~1}]\label{thm:bv}
Under no-arbitrage, the market value of an asset under book-value
wealth taxation is
\begin{equation}\label{eq:bv_oneperiod}
  V = \frac{1}{1 - \tw}\left(V^0 - \frac{\tw B}{1 + r_f}\right) ,
\end{equation}
where $V^0$ is the pre-tax market value, $B$ is the book value, and
$r_f$ is the risk-free rate.
\end{theorem}

\noindent
When the book-to-market ratio $\theta = B/V^0 < 1$ (the empirically
dominant case), the asset is worth \emph{more} under book-value
taxation than under market-value taxation: the investor pays tax on a
smaller base.  For a typical Norwegian unlisted holding with
$\theta \approx 0.4$, the effective assessment relative to market
wealth is $\beta\,\theta \approx 0.32$---the tax base is roughly a
third of market value.

\subsection{Pareto tail and demographic turnover}\label{sec:fp_pareto}

Pure GBM produces a spreading Gaussian; a stationary distribution
requires an additional mechanism.  Following \citet{Gabaix2009},
demographic turnover at rate $\delta > 0$ (each individual replaced by
a new entrant near the mean) adds a source-sink term:
\begin{equation}\label{eq:fp_turnover}
  \frac{\partial \pi}{\partial t}
  = -v\,\frac{\partial \pi}{\partial x}
    + D\,\frac{\partial^2 \pi}{\partial x^2}
    - \delta\,\pi + \delta\,\phi(x) \,,
\end{equation}
where $\phi(x)$ is the entrant distribution.  The stationary right tail
$\pi_{\mathrm{ss}}(x) \propto e^{-\zeta x}$ satisfies the
characteristic equation $D\zeta^2 - v\zeta - \delta = 0$, with
positive root:
\begin{equation}\label{eq:pareto}
  \zeta = \frac{v + \sqrt{v^2 + 4D\delta}}{2D} \,.
\end{equation}
Since $\pi_{\mathrm{ss}} \propto e^{-\zeta x}$ in log-wealth
corresponds to $p(W) \propto W^{-(1+\zeta)}$ in wealth, this is a
Pareto distribution with tail exponent~$\zeta$.  For representative
Norwegian parameters ($\sigma \approx 0.30$, $\mu \approx 0.08$,
$\delta \approx 1/30$), this gives $\zeta \approx 1.5$--$2.0$.

The Gini coefficient for a pure Pareto tail is
$\Gini = 1/(2\zeta - 1)$; for $\zeta = 1.5$, this gives
$\Gini = 0.5$.  Increasing the wealth tax raises $\zeta$ (steeper
tail, less inequality) through the drift-shift
$v_\tau = v - \tw < v$.

\begin{remark}[Tail stability under heterogeneous returns]
Equation~\eqref{eq:pareto} derives $\zeta$ from a homogeneous model
in which all agents share the same drift~$v$ and diffusion~$D$.
\citet{BernardBouchaudLeDoussal2026} solve the mean-field model with
quenched heterogeneous growth rates and show that the tail exponent
in the partially localised phase takes the form
$\mu = 1 - \Sigma_0^2/\sigma^4$, where $\Sigma_0^2$ is the
cross-sectional variance of growth rates.  The exponent thus depends
on the \emph{dispersion} of returns, not just the mean drift.
This is consistent with the empirical finding
(\Cref{sec:pareto_empirical}) that the Hill estimate
$\hat{\alpha} \approx 1.3$ is stable across all years of the
Kapital~400 panel, including the post-reform period: a uniform
drift shift (wealth tax increase) changes~$v$ but does not
affect~$\Sigma_0^2$, so the tail exponent is approximately invariant
under the tax change.  The stability of~$\hat{\alpha}$ is therefore
a prediction of the heterogeneous-return model, not a coincidence.
\end{remark}

The spectral gap of the Fokker--Planck operator determines the
relaxation timescale---how quickly the wealth distribution converges to
its new steady state after a policy change.  For realistic parameters:
\begin{equation}\label{eq:halflife}
  t_{1/2} = \frac{\ln 2}{\Lambda}
  \approx 21\;\text{years} \,.
\end{equation}

\subsection{Migration as a permeable boundary}\label{sec:fp_migration}

The taxonomy of Fokker--Planck modifications in \citet{Froeseth2026R}
classifies migration as a \emph{permeable boundary}: the confining
potential that sustains the steady state operates only up to a wealth
threshold beyond which agents exit the jurisdiction.  The reflecting
boundary at infinity is replaced by a partially absorbing boundary:
\begin{equation}\label{eq:migration}
  J(x_m, t) = \gamma\,\pi(x_m, t) \,,
\end{equation}
where $J$ is the probability current and $\gamma > 0$ is a migration
rate.  The consequence is a truncated Pareto tail: the right tail is
cut not by policy design but by agent exit, and the effective Gini
reduction is smaller than the closed-boundary prediction.

At the macroscopic level, the permeable-boundary picture rationalises
the quasi-binary regional response discussed in \Cref{sec:tipping}:
once the push rate exceeds the effective migration barrier, the
outflow is large and discrete rather than a smooth function of the
local tax differential.  Small interregional differentials generate
negligible boundary flux, while a discrete policy gap opens the
boundary.

\begin{remark}[Micro-foundation of the scalar ODE]\label{sec:fp_reduction}
The scalar ODE~\eqref{eq:main} arises from a coupled
Fokker--Planck system for resident and emigrant densities
$\pi_R(x,t)$, $\pi_E(x,t)$, with contagion-dependent transfer rate
$\lambda(\tilde{n},t)$ and return rate~$\rho$.  Integrating the
emigrant equation over log-wealth recovers~\eqref{eq:main} exactly
when $\lambda$ is wealth-independent, and approximately under a
separation of time scales: wealth distributions equilibrate on a time
scale ${\sim}\,D^{-1}$, while migration cascades unfold over years.
\end{remark}

\section{Empirical evidence from the Kapital~400}\label{sec:data}
\subsection{Data and timing conventions}\label{sec:timing}

\begin{figure}[tp]
  \centering
  \includegraphics[width=\textwidth]{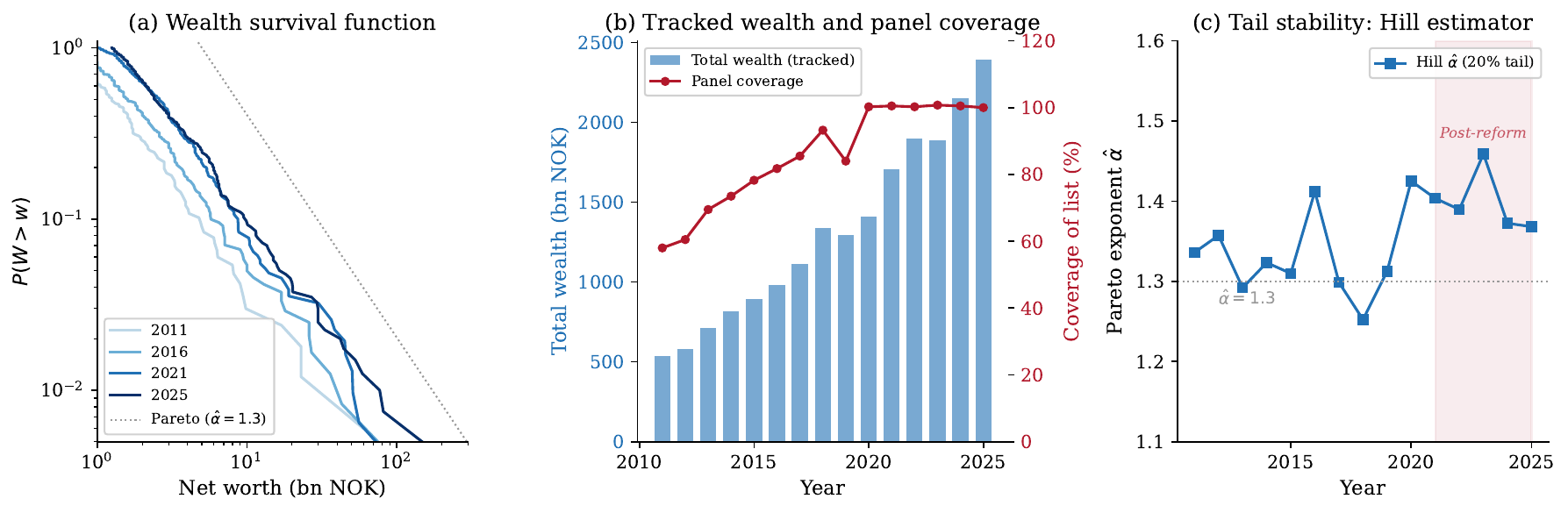}
  \caption{Wealth distribution of the Kapital~400 panel (2011--2025).
    \textbf{(a)}~Complementary CDF (survival function) on log-log
    axes for selected years.  The approximate linearity confirms the
    Pareto tail; the dashed reference line corresponds to
    $\hat{\alpha} = 1.3$.
    \textbf{(b)}~Total tracked wealth (bars) and panel coverage as a
    fraction of the published 400-entry list (circles, right axis).
    Coverage is incomplete before $\sim$2020 because the
    Kapital~Index database retains historical data only for persons
    who appear on the current list; those who fell below the entry
    threshold are removed.  Total wealth for earlier years is
    therefore a lower bound.
    \textbf{(c)}~Hill estimator of the Pareto exponent at the 20\%
    tail fraction.  The exponent is remarkably stable across the
    entire sample period, including the post-reform years (shaded),
    indicating that the tail shape is unaffected by the emigration
    episode.}
  \label{fig:wealth_distribution}
\end{figure}

Kapital, a business magazine published by Hegnar Media,%
\footnote{Hegnar Media also publishes the financial newspaper
Finansavisen; the Kapital~400 data are hosted on the shared web
platform at \texttt{finansavisen.no/kapital}.}
publishes an annual list of the 400 wealthiest
Norwegians each autumn \citep{Kapital400}.%
\footnote{Data were collected from individual person pages on
\texttt{finansavisen.no/kapital-index}, which report market-wealth
estimates, taxable assets, and tax paid for each list member.  The
panel was constructed by scraping individual person pages
incrementally across multiple list years and merging with the current
(2025) edition.  The resulting dataset contains 569 unique persons:
all 400 on the current list plus 169 who appeared in earlier editions
but have since dropped off (due to death, wealth falling below the
entry threshold, or editorial reclassification).  The online
Kapital~Index retains historical data only for persons who appear on
the current list, so the historical coverage of the full 400-entry
cohort is incomplete for early years---42\% of the 2011 cohort
is missing---but reaches full coverage by 2020
(see \Cref{fig:wealth_distribution}b).}
We construct a panel of 569 unique persons
over 2011--2025.  For each person-year
the data include an estimate of market wealth (in NOK billions), net
taxable assets, and tax paid.

The timing structure is important.  The market-value estimate targets
1~September of the publication year~$t$ (the list is typically
released later that month).  Taxable assets and tax paid come from the
public filing for tax year $t - 1$.  Within taxable assets, two
distinct lags apply:
\begin{enumerate}[nosep]
  \item For directly held listed shares, the assessment base is the 1
    January year-$t$ price---an $\sim$8-month lag relative to the
    market-value estimate.
  \item For unlisted shares, the assessment base is the statutory book
    value (\emph{skattemessig formuesverdi}) at the beginning of tax
    year $t - 1$---a $\sim$20-month lag.
\end{enumerate}
Unlisted shares are further subject to the statutory assessment
fraction $\beta = 0.80$ applied to book value.  For a typical holding
company with book-to-market ratio $\theta \approx 0.4$, the effective
assessment relative to market wealth is $\beta\,\theta \approx 0.32$:
the tax base is roughly a third of market wealth
(\Cref{thm:bv}; \citealp{Froeseth2026N}, \S9).

The \emph{ligningsformue} (taxable net wealth) column adds a further complication.  Kapital's
market-value column aggregates holdings under the family head, whereas
ligningsformue refers to the \emph{individual} tax filing.  When
shares have been restructured into children's names or holding
companies, ligningsformue may collapse without any change in economic
control.  A prominent case from the dataset: a fish-farming heir
formally inherited the bulk of the family's listed-company stake in
2013, collapsing the father's ligningsformue, but the father retained
voting control via A-shares and Kapital continues to list the combined
family wealth under his name.

\subsection{The Pareto tail}\label{sec:pareto_empirical}

A Pareto-tailed distribution is characterised by a power-law survival
function (complementary CDF):
\begin{equation}\label{eq:pareto_cdf}
  P(W > w) \;\propto\; w^{-\alpha}, \qquad w \to \infty \,,
\end{equation}
which appears as a straight line with slope $-\alpha$ on log-log axes.
We estimate the tail exponent directly from the market-wealth column
using the Hill estimator:
\begin{equation}\label{eq:hill}
  \hat{\alpha}_k = \frac{k}{\sum_{i=1}^{k}
    \bigl[\ln W_{(i)} - \ln W_{(k+1)}\bigr]} \,,
\end{equation}
where $W_{(1)} \geq \cdots \geq W_{(n)}$ are the order statistics and
$k$ is the number of upper-tail observations.  The results for 2025
($n = 400$, $W_{\max} = 262$~bn, $W_{\min} = 1.25$~bn):
\[
  \hat{\alpha}_{10\%} = 1.21, \qquad
  \hat{\alpha}_{20\%} = 1.37, \qquad
  \hat{\alpha}_{30\%} = 1.20 \,.
\]
Across all years the Hill exponent is remarkably stable, averaging
$\hat{\alpha} \approx 1.30$ at the 20\% tail fraction with no
discernible time trend.  This is consistent with---but somewhat
below---the theoretical prediction $\zeta \approx 1.5$--$2.0$ from
\eqref{eq:pareto}, as expected for a sample that conditions on the top
400 and therefore truncates the body of the distribution.

Concentration measures confirm the fat tail: the top~10 fortunes hold
29--37\% of total Kapital~400 wealth depending on year, and the
within-cohort Gini ranges from 0.54 to 0.60.

\subsection{Identifying emigrants}\label{sec:emigrants}

Cross-referencing the panel with biographical descriptions on each
person page and the ``emigration fingerprint'' (tax paid collapsing to
near-zero while market wealth continues) yields 67 direct emigrants
among current and former Kapital~400 members; a separate
investigation of intergenerational wealth transfers identifies an
additional 36 recent heir-emigrants (\Cref{sec:heir_emigrants}).\footnote{The
underlying data are from publicly available Kapital~400 person pages.
To protect the privacy of living individuals, the seven
Blandhol-window emigrants are identified by coded labels rather than
by name.  The well-known long-term expatriates and post-reform
departures are public figures whose emigration is widely reported.
Heir-emigrants were identified from Finansavisen coverage and the
Kapital~300 heir list, and cross-verified against Kapital~400 family
entries.  An additional six heirs classified as long-term foreign
residents (abroad well before 2016) are excluded from the recent
count.}

At least 16 are long-term expatriates (pre-2016): Fredriksen (1978),
Hagen (1968), Siem (1972), Smedvig (1991),
Stolt-Nielsen (2001), Tr{\o}im (1995), among others.  The high-profile
post-reform departures---R{\o}kke (Sep 2022), Klaveness (2022),
a prominent shipping-family estate division (2021), Moan (2024)---all occurred \emph{after}
Blandhol's 2016--2020 window closes.

Excluding the long-term expatriates, 27 direct emigrants departed
during the policy-relevant window 2016--2025, carrying a combined
market wealth of approximately 172~bn~NOK\@.  These 27 are the focus
of the analysis below.

Seven of them emigrated within the Blandhol window (2016--2020).
\Cref{tab:emigrants} characterises each by sector, wealth bracket,
emigration year, and the nature of their wealth (active control versus
passive holdings).

\begin{table}[t]
\centering
\caption{Kapital~400 emigrants within the Blandhol sample window
(2016--2020).  Market wealth and ligningsformue (lig.) are in NOK
billions as of the last year the individual appeared on the domestic
tax rolls.}\label{tab:emigrants}
\renewcommand{\arraystretch}{1.3}
\begin{tabular}{@{}llrrlp{4.2cm}@{}}
\toprule
\textbf{ID} & \textbf{Sector} & \textbf{Wealth} & \textbf{Lig.} &
  \textbf{Year} & \textbf{Type} \\
\midrule
E1 & Financial services & 9.0 & 7.24 &
  2018 & Passive (sole owner of financial holding) \\
E2 & Diversified portfolio & 8.7 & 5.10 &
  $\sim$2020 & Passive (heir; professional CEO) \\
E3 & Commercial real estate & 3.0 & 0.21 &
  $\sim$2019 & Active (operating business) \\
E4 & Technology (post-exit) & $\sim$1 & --- &
  $\sim$2016 & Passive (post-liquidity event) \\
E5 & Technology (post-exit) & $\sim$1 & --- &
  $\sim$2016 & Passive (post-liquidity event) \\
E6 & Technology (post-exit) & $\sim$1 & --- &
  $\sim$2016 & Passive (post-liquidity event) \\
E7 & Financial services & 1.1 & --- &
  $\sim$2017 & Passive (sold stake; now investor) \\
\bottomrule
\end{tabular}
\end{table}

Of these seven, only E3 ran a domestic operating business.  The rest
are financial investors, passive heirs, or post-exit entrepreneurs.
Moreover, the departures are not independent events: E1 and E7 were
former business partners; E4--E6 moved together after a shared
liquidity event.  E2's sibling remains in Norway and continues to pay
tax at full rates; Kapital lists the combined family wealth under
both siblings separately but with identical totals.

\subsection{The hidden channel: heir-emigrants}\label{sec:heir_emigrants}

The 27 direct emigrants identified above are persons who themselves
appear on the Kapital~400 list and whose emigration is detectable
through the ligningsformue fingerprint or the profile residence field.
A second, quantitatively comparable channel is invisible in the panel
data: \emph{heir-emigration}, in which the controlling owner transfers
economic ownership (B-shares or equivalent non-voting equity) to
children or grandchildren who then emigrate, while the controlling owner
retains A-shares (voting control) and remains a Norwegian tax
resident.

The Kapital~400 lists family wealth under the controlling owner's name.  When
an heir emigrates with B-shares, the controlling owner's ligningsformue may
collapse---but the list attributes the market wealth to the family
unit, not to the individual heir.  The heir never appears as a
separate entry and is therefore undetectable through within-panel
methods.

Systematic cross-referencing of Kapital~400 family profiles,
Kapital~300 heir profiles, and Norwegian business press coverage
identifies 36 recent heir-emigrants carrying a combined wealth of
approximately 127~bn~NOK\@.\footnote{Primary sources: Finansavisen,
``Stadig flere unge velger Sveits,'' 28~August 2025 (22 Swiss
heir-emigrants); and the Kapital~300 heir list
(\url{https://www.finansavisen.no/kapital-index/norges-rikeste-arvinger}),
accessed October~2025, which adds 11 heirs not in the August tabulation
(including non-Swiss destinations).  All entries were cross-verified
against the Kapital~400 panel by controlling owner name and person identifier.
An additional six heirs are classified as long-term foreign residents
(abroad well before any tax-policy change) and one pre-2016 emigrant
(2009); these are excluded from the recent count.}
\Cref{tab:heir_emigrants} reports the largest cases.

\begin{table}[tp]
\centering
\caption{Largest identified recent heir-emigrants from Kapital~400
  and Kapital~300 families.  ``Wealth'' is the journalist-estimated
  economic value held by the heir (Kapital~300 estimate, October~2025).
  All figures in NOK billions.  The table shows the 12 largest cases;
  24 additional heirs carry a combined 24.6~bn.  Six long-term foreign
  residents (20.0~bn) and one pre-2016 emigrant (21.0~bn)
  are excluded.  To protect the privacy of heirs---who are typically
  in their 20s--30s and not public figures---entries are coded
  analogously to \Cref{tab:emigrants}.}\label{tab:heir_emigrants}
\footnotesize
\renewcommand{\arraystretch}{1.2}
\begin{tabular}{@{}lrrrlp{3.2cm}@{}}
\toprule
\textbf{ID} & \textbf{Heirs} & \textbf{Wealth} &
  \textbf{Year} & \textbf{Dest.} & \textbf{Sector} \\
\midrule
H1  & 2 & 25.6 & 2022 & CH & Grocery / diversified \\
H2  & 1 & 14.5 & 2022 & CH & Industry / diversified \\
H3  & 1 & 14.2 & 2022 & CH & Real estate \\
H4  & 1 &  7.6 & 2023 & UK & Shipping / finance \\
H5  & 1 &  7.4 & 2025 & SE & Real estate \\
H6  & 1 &  6.5 &  --- & CH & Real estate \\
H7  & 2 &  6.6 & 2022 & IT & Industry / diversified \\
H8  & 1 &  4.8 & 2021 & UK & Shipping \\
H9  & 2 &  4.7 & 2024--25 & CH/IT & Shipping \\
H10 & 1 &  3.6 &  --- & CH & Real estate \\
H11 & 1 &  3.5 &  --- & CH & Consumer goods \\
H12 & 1 &  3.1 &  --- & CH & Technology \\
\midrule
\multicolumn{3}{@{}l}{\textit{24 additional heirs}} & & & \\
\midrule
\textbf{Total} & \textbf{36} & \textbf{126.6} & & & \\
\bottomrule
\end{tabular}
\end{table}

Several features of the heir-emigrant channel are noteworthy.  First,
the scale is comparable to direct emigration: 127~bn~NOK in recent
heir wealth versus 172~bn among the 27 direct emigrants, giving a
combined total approaching 300~bn~NOK\@.  Heirs account for 57\% of
emigrant persons and 42\% of emigrant wealth.  Second, Switzerland
remains the dominant destination (25 of 36 recent heirs, 69\%), but
the heir channel shows more destination diversity than direct
emigration: three heirs moved to the UK, three to Italy, two to
Denmark, and one each to Sweden, Cyprus, and the Netherlands.  Third,
the mechanism is specifically tied to the Norwegian dual share-class
structure and the \emph{fritaksmetoden}: by transferring B-shares
(which carry economic value but no voting rights) while retaining
A-shares, the controlling owner can engineer a generational transfer that
simultaneously achieves succession planning and tax-base
migration---without any operational disruption to the underlying
enterprise.

Six additional heirs---carrying a combined 20~bn~NOK---are classified
as long-term foreign residents who have lived abroad well before any
tax-policy change and are excluded from the recent count.  Several
were born abroad or moved as children and never resided in Norway as
adults.  The distinction matters: the ``utflyttet'' flag in Kapital's
data does not distinguish recent tax-motivated emigrants from heirs
who never lived in Norway.

The productivity haircut for heir-emigrants is, by construction, zero:
the heir held no management role, the controlling owner continues to
run the company, and the professional management team remains in
Norway.
This is the strongest empirical evidence for the wealth--control
separation (F5 in \Cref{sec:facts}): the emigrating wealth is pure
economic exposure with no attached human capital.

Why is the heir channel so much more active than the direct channel?
\citet{FriedmanEtAl2025}, drawing on in-depth interviews with 35
individuals in the top~1\% of the UK wealth distribution, find that
place-specific attachments---career networks, family proximity, and
above all the cultural infrastructure of London---overwhelmingly
dominate tax considerations in location decisions.  None of their
interviewees were planning to move for tax reasons; many drew sharp
moral boundaries against those who do, dismissing low-tax destinations
as ``boring and culturally barren.''  These place-specific anchors
accumulate over a lifetime and are strongest for established
senior owners in their 60s and 70s.  Heirs in their 20s and 30s, by
contrast, have not yet built comparable location-specific capital:
they have weaker professional attachment, thinner social networks, and
face lower stigma for relocating (they are ``starting a life abroad''
rather than ``fleeing taxes'').  The generational asymmetry in
place-specific capital thus explains why the intergenerational
transfer moment is the critical juncture for tax-base migration.

\begin{remark}[The Great Wealth Transfer and heir-emigration]
The Norwegian heir-emigrant channel is likely an early instance of a
broader phenomenon.  Aggregate bequeathable wealth in the United
States rose from 256\% to 425\% of GDP between 1997 and 2021, with
97\% of the increase accruing to households aged~55 and older and
75\% to the top decile within that group \citep{GaleEtAl2024}.
Comparable demographic dynamics apply in Western Europe.  Each
intergenerational transfer creates a decision point at which the
heir can choose tax jurisdiction.  \citet{AdvaniEtAl2025} find that
among UK super-rich taxpayers, those not attached to the local labour
market respond significantly more strongly to tax changes---the
elasticity is concentrated among passive wealth-holders.  Combined
with the place-specific capital mechanism documented above
\citep{FriedmanEtAl2025}, heirs are the limiting case on both
dimensions: no operational role and no accumulated location-specific
capital, making them maximally mobile.

Countries that combine wealth or inheritance taxation with an ageing
ultra-wealthy cohort should therefore see heir-emigration accelerate
during the peak transfer years (roughly 2025--2045), particularly
where dual share-class structures allow economic and control rights to
be separated.
\end{remark}

\subsection{Descriptive patterns consistent with contagion}\label{sec:contagion_evidence}

The emigration episode exhibits four descriptive patterns that a
standard independent-response model cannot easily generate.  We stress
that these are suggestive, not causally identified; the data cannot
distinguish social contagion from common shocks or correlated
preferences.  Nonetheless, the conjunction of all four raises the
empirical cost of maintaining the independent-response null.

\paragraph{Reversed sequencing and destination convergence.}
The temporal structure is the reverse of what the standard model
predicts.  Phase~1 (2018--2021) comprises roughly a dozen direct
departures, led by a single financial-services principal in~2018 (E1) and
followed by a handful of real-estate and finance figures plus
a three-sibling shipping-family cluster (${\sim}60$~bn~NOK combined).  Phase~2
(2022--2024) adds roughly fifteen direct departures totalling
${\sim}90$~bn~NOK, plus numerous heir-emigrations
(H1--H12 and others,
together ${\sim}127$~bn across 36 identified recent heirs).
Including both channels, total emigrating wealth approaches
300~bn~NOK\@.
Emigration years are identified from the \emph{ligningsformue}
fingerprint: when a person's taxable wealth collapses relative to
their market valuation while remaining on the Kapital~400 list, they
have left the Norwegian tax base.  R\o{}kke's 2022 departure is the
clearest case: ligningsformue fell from 18.6~bn to 1.3~bn while net
worth remained above 35~bn, and the exit year coincided with a record
\emph{utlignet skatt} of 1013~mn~NOK\@.  Under independent response,
the largest fortunes---facing the highest absolute burden---should
move first; instead, a small vanguard of mid-fortune holders moved
before the largest followed, precisely the pattern generated by a
threshold model \citep{Granovetter1978}.  Simultaneously, pre-2016
emigrants dispersed across the UK, US, Monaco, Cyprus, and
Switzerland, whereas post-2021 departures converge overwhelmingly
on Swiss low-tax cantons (Zug, Schwyz, Lugano).  The heir-emigrant
data reinforce this: 25 of 36 recent heirs (69\%) moved to
Switzerland, though the heir channel shows somewhat more destination
diversity (UK, Italy, Scandinavia) than the direct channel.
A 2025 cross-section confirms the concentration: 33 of 66
foreign-resident Kapital~400 members reside in Switzerland.

\paragraph{Network and family clusters.}
The departures are not independent draws.  \Cref{tab:clusters}
catalogues the identifiable clusters among post-2016 emigrants,
including both direct emigrants and heir-emigrants
(\Cref{sec:heir_emigrants}).  The clustering by professional network,
family, and intergenerational transfer---shared advisors, direct
encouragement, reduction of perceived stigma through peer
example---exceeds what independent optimisation would predict.

\begin{table}[tp]
\centering
\caption{Network and family clusters among post-2016 Kapital~400
  emigrants.  $N$: number of identified members (direct $+$ heir).
  Link type characterises the social or informational channel
  connecting cluster members.  Heir-emigrant clusters (bottom panel)
  involve B-share transfers where the controlling owner remains in
  Norway.}\label{tab:clusters}
\footnotesize
\renewcommand{\arraystretch}{1.2}
\begin{tabular}{@{}p{2.8cm}cp{1.7cm}cp{2.8cm}p{3.0cm}@{}}
\toprule
\textbf{Cluster} & \textbf{$N$} & \textbf{Sector} & \textbf{Phase}
  & \textbf{Link type} & \textbf{Compound shock} \\
\midrule
\multicolumn{6}{@{}l}{\textit{Direct emigrants}} \\[2pt]
Tech startup (E4--E6)
  & 3 & Technology & 1
  & Colleagues; shared exit
  & Post-liquidity event \\
Fintech co-founders
  & 2 & Tech / finance & 2
  & Co-founders
  & Post-liquidity event \\
Finance network (E1, E7)
  & 2 & Finance & 1
  & Former partners
  & --- \\
Real estate milieu
  & ${\sim}10$ & Commercial RE & 1--2
  & Shared advisors, brokers, industry forums
  & Valuation tightening $+$ rate reversal $+$ generational transition \\
Shipping/cruise family heirs
  & 3 & Hotels / cruise & 1
  & Family (estate division)
  & --- \\
Aquaculture
  & 2 & Salmon farming & 2
  & Sector peers
  & \emph{Grunnrenteskatt} $+$ production tax $+$ wealth tax \\
High-profile individuals
  & ${\sim}6$ & Mixed & 2
  & ---
  & --- \\
\midrule
\multicolumn{6}{@{}l}{\textit{Heir-emigrants (B-share transfers, controlling owner remains in Norway)}} \\[2pt]
Mega heirs ($>$10~bn)
  & 4 & Diversified & 2
  & Intergenerational B-share transfer
  & Wealth tax $+$ exit-tax anticipation \\
Mid-tier heirs (1--10~bn)
  & ${\sim}20$ & Mixed & 1--2
  & Intergenerational transfer; Swiss advisory network; UK/IT lifestyle
  & --- \\
Sub-billion heirs
  & ${\sim}12$ & Mixed & 2
  & Intergenerational transfer
  & --- \\
\bottomrule
\end{tabular}
\end{table}

\begin{figure}[tp]
  \centering
  \includegraphics[width=\textwidth]{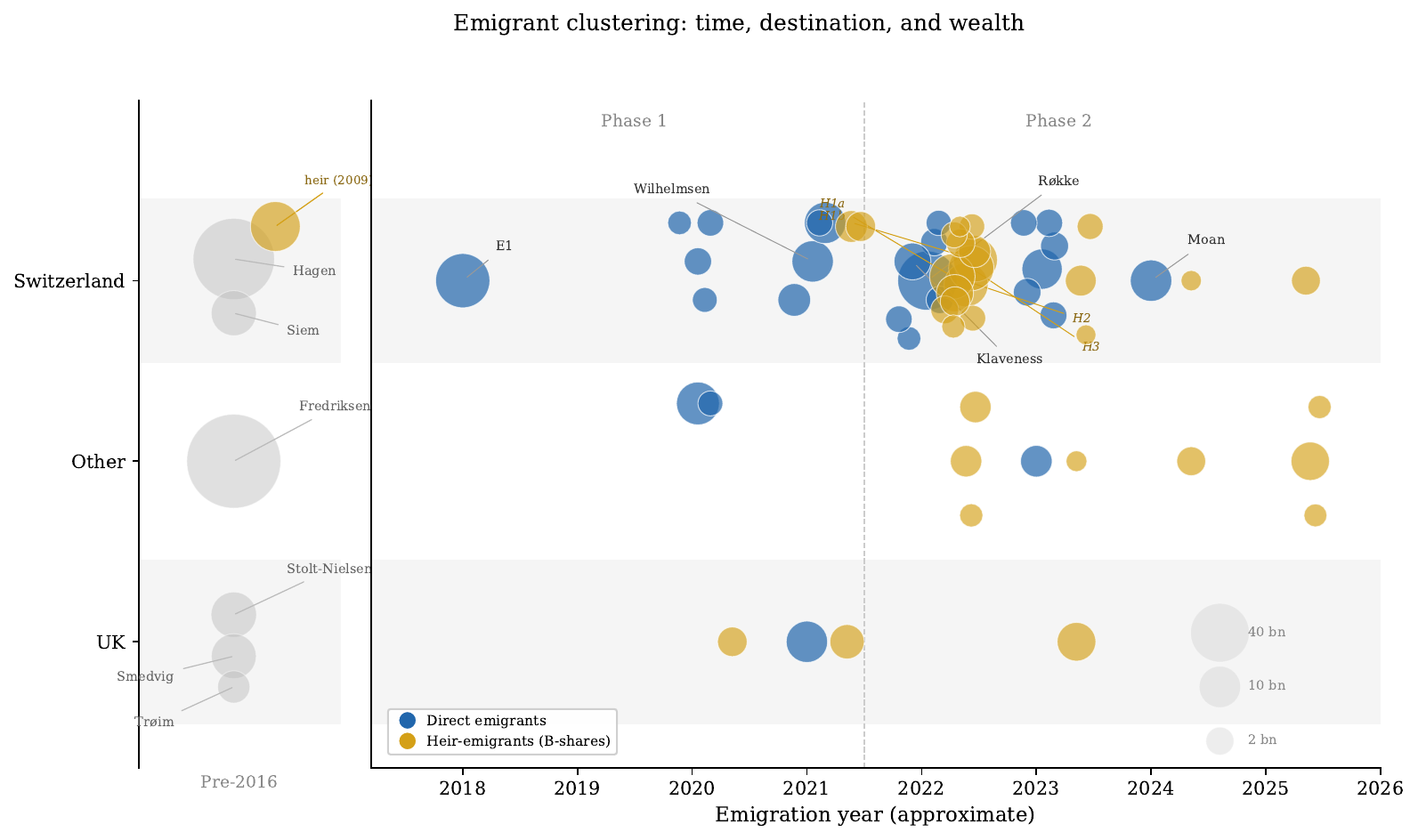}
  \caption{Emigrant clustering in time, destination, and wealth.
    Blue circles represent direct emigrants; gold circles represent
    heir-emigrants (B-share transfers where the controlling owner remains in
    Norway).  Bubble area is proportional to market wealth (NOK~bn).
    The left margin shows pre-2016 long-term expatriates for scale;
    the main panel covers 2018--2025.  The dashed line separates
    Phase~1 (small vanguard) from Phase~2.  Only widely reported
    public-figure departures are labelled by name; other emigrants are
    identified by coded labels (\Cref{tab:emigrants,tab:heir_emigrants})
    or left unlabelled.  Emigration years for direct emigrants are
    inferred from the \emph{ligningsformue} fingerprint; heir-emigrant
    years are known for approximately ten cases from press sources and
    estimated for the remainder.  The vertical axis illustrates the
    convergence from dispersed pre-2016 jurisdictions to near-complete
    Swiss concentration.}
  \label{fig:emigrant_clusters}
\end{figure}

\paragraph{Sector-specific compound shocks.}
The sector composition reflects compound shocks beyond the wealth tax
alone.  Real-estate investors span both phases: assessment
rules tightened sharply (secondary housing discounts removed;
\emph{gjeldsreduksjon} crystallising positive tax bases on previously
negative-net-wealth portfolios; \citealp{Froeseth2026E}, \S9.3), and
the interest-rate reversal (0\% in May~2020 to 4.50\% by
December~2023) hit leveraged property investors hardest.  For
founder-operators aged 60--65 who built portfolios during decades of
falling rates, the simultaneous reversal coincided with natural
succession questions.  These conventional channels explain the
sectoral \emph{composition}; what they do not explain is the
\emph{clustering} in time and destination among operators in different
cities and market segments who share brokers, bankers, and industry
forums.  In aquaculture, a layered shock---traffic-light system
(2017), per-kilogram production tax (2021), 25\% resource rent tax
passed retroactively in 2023 after a 40\% proposal that caused listed
stocks to fall 15--19\% overnight---far exceeds the wealth tax alone.
In both sectors, the compound shock provided the \emph{motive}; the
network shaped the \emph{timing} and \emph{destination}.

\paragraph{The micro-elasticity gap.}
\citet{JakobsenEtAl2024} estimate that a 1\,pp wealth tax increase
reduces the stock of wealthy residents by ${\sim}2\%$.  Applied to
Norway (effective rate increase ${\sim}0.25$\,pp on listed wealth),
this predicts ${\sim}15$ additional departures from a population of
${\sim}3{,}000$.\footnote{The calculation applies the Jakobsen
et~al.\ elasticity to the Norwegian wealth-tax-paying population
above 10\,M~NOK\@.  The compound shock also included increases in
dividend and capital-gains taxation, which complicates the
attribution.}  Observed departures---254 in 2022 and 261 in 2023
among those above 10\,M~NOK---exceed this by an order of magnitude,
consistent with an amplification mechanism.

\section{The micro-to-macro gap}\label{sec:gap}

The aggregate output effect of emigration depends on the productivity
haircut.  To see how, start from a Cobb--Douglas aggregate production
function $Y = A\,K^{\alpha_K} L^{1-\alpha_K}$, where $K$ is the
domestic capital stock.  When an emigrant~$i$ departs, the
domestically productive fraction of her capital falls from $K_i$ to
$K_i\,e^{-\eta_i}$, where $\eta_i \geq 0$ is the productivity haircut
(the proportional reduction in capital that is effectively deployed
domestically after emigration).  Summing over all emigrants and
log-differencing:
\begin{equation}\label{eq:output_general}
  \frac{\Delta Y}{Y}
  \approx \alpha_K \sum_{i \in E}
    (1 - e^{-\eta_i})\,\frac{K_i}{K}
  = \alpha_K \sum_{i \in E}
    (1 - e^{-\eta_i})\,\omega_i \,,
\end{equation}
where $\omega_i = K_i / K$ is emigrant~$i$'s share of the domestic
capital stock.  With the two-type decomposition of \Cref{sec:types}
(active-control owners~$\mathcal{A}$ with haircut~$\eta_A$ and passive
holders~$\mathcal{P}$ with haircut~$\eta_P \approx 0$), this becomes:
\begin{equation}\label{eq:output}
  \frac{\Delta Y}{Y}
  \approx \alpha_K\,\Bigl[
    (1 - e^{-\eta_A})\,\omega_E^{(\mathcal{A})}
    + \underbrace{(1 - e^{-\eta_P})}_{\approx\, 0}\,
      \omega_E^{(\mathcal{P})}
    \Bigr] \,,
\end{equation}
where $\omega_E^{(\mathcal{A})}$ and $\omega_E^{(\mathcal{P})}$ are
the aggregate wealth shares emigrated by each type.  The key
implication is immediate: when most emigrating wealth is passive
($\omega_E^{(\mathcal{P})} \gg \omega_E^{(\mathcal{A})}$) and
$\eta_P \approx 0$, the output effect is negligible regardless of
the total emigrating wealth.

\citet{Blandhol2025} estimates a 12.6\% revenue decline and scales it
to a 1.3\% GDP loss.  The path from the event-study estimate to this
aggregate figure passes through five identification conditions, each
of which fails.

\subsection{Market wealth versus taxable wealth}\label{sec:valuation}

The Fokker--Planck framework (\Cref{sec:fp}) models the evolution of
\emph{market} wealth $x = \ln W$.  Blandhol's treatment population is
defined by \emph{taxable} wealth exceeding 100M NOK\@.  The mapping
between the two is mediated by the book-value assessment and the
statutory assessment fraction described in \Cref{sec:timing}: for a
typical unlisted holding with $\theta = 0.4$ and $\beta = 0.80$, the
effective assessment is 32\% of market value (\Cref{thm:bv}).
Blandhol's threshold of 100M NOK in taxable wealth therefore
corresponds to roughly 250--400M NOK in market wealth, depending on
portfolio composition.

The composition of these distortions is nonlinear,
asset-class-dependent, and itself a symmetry-breaking channel
(\Cref{sec:drift_shift}, \Cref{eq:ch1}): it creates anisotropic
effective drift across asset classes, distorting portfolio choice.
The GDP integral below runs over market wealth, not taxable wealth;
Blandhol estimates a revenue effect in taxable-wealth space and applies
it to a GDP calculation in market-wealth space, with the nonlinear
mapping between the two unaccounted for.

\subsection{Wealth-weighted representativeness}\label{sec:representativeness}

The aggregate output effect can be written as an integral over the
emigrant wealth distribution:
\begin{equation}\label{eq:output_integral}
  \frac{\Delta Y}{Y}
  = \alpha_K \int_{x_{\min}}^{\infty}
    \bigl(1 - e^{-\eta(x)}\bigr)\;
    \omega_E(x)\;\dd x \,,
\end{equation}
where $\omega_E(x) = e^x\,\pi_E(x)\big/\int e^x\,\pi_E\,\dd x$ is
the wealth-weighted emigrant density and $\eta(x)$ is the productivity
haircut at log-wealth level~$x$.  This is the continuous-wealth
generalisation of~\eqref{eq:output_general}.

Blandhol's event study estimates an average revenue effect
$\hat{\mu}^{\mathrm{DiD}}$ from a sample of $N \approx 5$ emigration
events with treated firms in a narrow band of the taxable wealth
distribution.  For this estimate to deliver a valid GDP figure, the
sample must satisfy a \emph{wealth-weighted representativeness
condition}: the wealth-weighted mean of the sampled emigrant
distribution must approximate that of the population,
\begin{equation}\label{eq:representativeness}
  \frac{1}{N}\sum_{i=1}^{N} W_i\,\eta_i
  \;\approx\;
  \EE_{\omega_E}\!\bigl[\eta(x)\bigr]
  = \int \eta(x)\,\omega_E(x)\,\dd x \,.
\end{equation}

The Pareto structure makes this condition extremely demanding.  Under a
Pareto distribution with exponent $\zeta < 2$, the wealth-weighted
measure $\omega_E(x)$ is even more top-heavy than the count measure
$\pi_E(x)$: weighting by $e^x$ shifts one unit of tail mass, giving an
effective exponent $\zeta - 1 < 1$ in the wealth-weighted density.
With $\hat{\alpha} \approx 1.3$ (\Cref{sec:pareto_empirical}), the
wealth-weighted exponent is $\approx 0.3$.

\begin{proposition}[Tail dominance]\label{prop:tail}
Let the emigrant wealth distribution follow a Pareto law with tail
exponent $\zeta$.  Then for any productivity haircut $\eta(x) \geq 0$
bounded away from zero on the upper tail:
\begin{enumerate}[nosep]
  \item The wealth-weighted expectation
    $\EE_{\omega_E}[\eta]$ is determined almost entirely by the
    behaviour of $\eta(x)$ for large~$x$.
  \item If $\zeta < 2$ (empirically $\hat{\alpha} \approx 1.3$), the
    variance of $\hat{\eta}$ estimated from $N$ i.i.d.\ draws from
    the body of the distribution diverges: the sample mean is an
    inconsistent estimator of the wealth-weighted population mean.
  \item For a sample of size $N$ drawn from wealth levels below the
    $p$-th percentile, the fraction of the wealth-weighted integral
    captured is at most $p^{(\zeta-1)/\zeta}$, which for
    $\zeta = 1.3$ and $p = 0.95$ gives $0.95^{0.23} \approx 0.99$
    but for $p = 0.50$ gives $0.50^{0.23} \approx 0.85$.  The top
    5\% of emigrants by wealth carry $>15\%$ of the integral's
    mass; the top 1\% carry $>5\%$.
\end{enumerate}
\end{proposition}

In the Kapital~400 data, the top 10 emigrant fortunes (Fredriksen,
R{\o}kke, etc.) hold more market wealth than
the remaining 57 identified direct emigrants combined.  None of them
emigrated during 2016--2020 with an active domestic operating company.
Adding the 36 recent heir-emigrants (\Cref{sec:heir_emigrants})
reinforces the point: 127~bn~NOK of emigrating wealth is, by
construction, entirely passive---the heir held no management role and
the controlling owner remained in Norway.  The wealth-weighted
integral is therefore dominated by individuals who are either entirely
absent from Blandhol's sample (pre-2016 and post-2020 emigrants) or
whose emigration carried negligible productivity implications (passive
heirs, financial investors, heir-emigrants).

The treatment population---households with net taxable wealth above
100M NOK---comprises roughly 1\,000--2\,000 households.  At an
emigration rate of ${\sim}0.2\%$ per year, and with only 41\% being
active firm owners \citep[Table~1]{Blandhol2025}, the firm event study
rests on $\sim\!4$--$8$ treated emigration events over 2016--2020.
The individuals driving the 12.6\% revenue estimate are therefore
likely mid-tier wealthy (100M--1B NOK in taxable wealth), whose firms
are small enough that individual industry shocks---an oil price drop
for an energy services firm, an interest rate hike for a real estate
developer---can dominate the estimate.

The B{\o} municipality experiment introduced in \Cref{sec:facts} is a
complementary within-Norway data point but does not by itself speak
to the wealth-weighted representativeness problem.  It is drawn from
a completely different sampling frame than Blandhol's firm event
study, and representativeness is a property of the
\emph{wealth-weighted} emigrant distribution: no local experiment
on thousands of taxpayers can resolve the tail contribution of the
handful of top fortunes that dominate the output integral.

\subsection{Signal-to-noise in short windows}\label{sec:signal_noise}

Even if the sample were cross-sectionally representative, the
time-domain coverage is insufficient.

Each firm revenue path $R_i(t)$ is an observation of a stochastic
process with both a drift component (the structural productivity
effect~$\eta$) and a diffusion component (idiosyncratic and industry
shocks).  For a single path observed over horizon~$T$, the
signal-to-noise ratio for extracting $\eta$ scales as
\begin{equation}\label{eq:snr}
  \mathrm{SNR}_i = \frac{|\eta_i|\,\sqrt{T}}{\sigma_R} \,,
\end{equation}
where $\sigma_R$ is the annual firm-level revenue volatility.  For
small and mid-size firms, $\sigma_R \in [0.15, 0.30]$.  With
$|\eta| \approx 0.12$ (the event-study point estimate) and $T = 5$:
\[
  \mathrm{SNR}_i \in [0.9,\, 1.8] \,.
\]
With $N \approx 5$ paths, the aggregate SNR is
$\sqrt{N}\,\mathrm{SNR}_i \approx 2$--$4$---marginal for inference,
and contingent on the paths being independent draws.  They are not:
all five paths share the same macro environment (oil price collapse
2014--2016, recovery 2017--2018, COVID 2020).

The Fokker--Planck relaxation analysis (\Cref{sec:fp_pareto}) provides
an independent perspective.  The half-life~\eqref{eq:halflife} of
$\sim\!21$ years means that Blandhol's 5-year window captures less
than one quarter of a single relaxation cycle.  Revenue fluctuations
over this horizon are dominated by the diffusion term
$\sigma_R\,\dd B_t$, not by the drift shift~$\eta$ that matters for
the long-run productivity effect.

\subsection{Revenue, wealth, and the control gap}\label{sec:revenue}

A final identification gap separates revenue from the wealth-path
variable in the Fokker--Planck framework.  The GDP
integral~\eqref{eq:output_integral} is indexed by market
wealth~$W$---the stock that enters the production function via
$Y = A\,K^\alpha\,L^{1-\alpha}$.  Blandhol's outcome variable is
firm \emph{revenue}~$R$, a flow measure related to capital through the
production function and to wealth through the ownership structure.

Under the wealth--control separation documented in
\Cref{sec:types,sec:heir_emigrants}, the emigrating party typically
holds economic exposure but no operational role.  Revenue responds
to \emph{control} decisions---investment, hiring, strategy---not to
the location of the economic owner.  The event-study revenue decline may
therefore reflect
\begin{enumerate}[nosep]
  \item a generational transition (the heir's emigration coincides
    with the parent's retirement and management handover),
  \item an industry shock correlated with the emigration decision
    (a commodity price decline that both reduces revenue and triggers
    the decision to move), or
  \item genuinely reduced oversight from the departing controller.
\end{enumerate}
Only (iii) corresponds to the productivity haircut~$\eta$ in the
Fokker--Planck framework.  Without decomposing the event-study sample
by share class (A-share versus B-share migrations), the three channels
are confounded.

Blandhol herself notes (p.~15) that the revenue loss could reflect
either emigration being costly for the firm or a negative productivity
shock driving both the revenue decline and the migration decision.
Her structural decomposition (p.~32) gives
$\mu^{\mathrm{DiD}} = \mu + \text{selection} +
\text{wealth accumulation}$,
but extracting $\mu$ requires the structural model to be correctly
specified---which assumes independent, rational emigration with no
social contagion, no reference dependence, and no distinction between
share classes.

\subsection{Summary of the identification gap}\label{sec:gap_summary}

The micro-to-macro extrapolation requires five conditions to hold
simultaneously:
\begin{enumerate}[nosep]
  \item The event-study sample is wealth-weighted representative of
    the emigrant population (violated: top missing,
    \Cref{sec:representativeness}).
  \item The observation window is long enough to separate structural
    drift from path noise (violated: $T \ll t_{1/2}$,
    \Cref{sec:signal_noise}).
  \item The outcome variable (revenue) identifies the productivity
    haircut~$\eta$ in the Fokker--Planck state space (violated:
    revenue confounds control and ownership effects,
    \Cref{sec:revenue}).
  \item The tax-wealth space in which the sample is selected maps
    cleanly to the market-wealth space over which the GDP integral is
    computed (violated: nonlinear book-value distortion,
    \Cref{sec:valuation}).
  \item The emigration response is identified as a response to the
    wealth tax, not confounded by simultaneous changes in the exit-tax
    regime or other tax channels (violated: the 2022 emigration wave
    coincided with the abolition of the five-year exit-tax lapse rule,
    and many emigrants held large deferred capital gains inside holding
    structures accumulated via the \emph{fritaksmetoden}).
\end{enumerate}
Each condition fails independently.  Together, they render the
1.3\%~GDP loss estimate uninformative about the true aggregate cost of
wealth-tax-induced emigration.

With a sample this small, drawn from a single historical episode during
a period of volatile commodity prices and rising interest rates, the
12.6\% estimate cannot credibly be separated from the specific
circumstances of the handful of individuals involved.  As
\citet{Popper1957} argued, generalising from a single historical path
requires strong structural assumptions---precisely the assumptions our
contagion model calls into question.

\section{Calibration}\label{sec:calibration}

The contagion model contains several parameters.  With 27 confirmed
direct departures and 36 recent heir-emigrants from Kapital~400/300
families
over a single policy episode, formal econometric estimation of the
contagion strength is not credible---and we should not pretend
otherwise.  (This is, after all, the same identification
problem we criticise in \Cref{sec:gap}: overparameterising a
structural model on a handful of observations from one historical
path.)  We therefore take a different approach: we identify the
parameters that the data \emph{can} discipline, report their values,
and are explicit about the parameters that remain unidentified.

\subsection{What the data identify}\label{sec:identified}

\paragraph{The visibility weight $\xi$.}
The ratio $\tilde{n}/n$ is a direct observable.  From \Cref{tab:emigrants}
and the Kapital~400 panel, the four tax-motivated emigration events
through 2022 (E1, a shipping-family estate division, R{\o}kke, Klaveness)
carried a combined market wealth of approximately 108~bn~NOK out
of a total pool of 1\,896~bn.  This gives
\begin{equation*}
  \frac{\tilde{n}(2022)}{n(2022)}
  = \frac{0.057}{0.009}
  \approx 6.3 \,.
\end{equation*}
For the empirical wealth distribution of the Kapital~400 (which has
Pareto exponent $\hat{\alpha} \approx 1.3$), the ratio
$\tilde{n}/n$ as a function of~$\xi$ can be computed exactly.
Matching the observed ratio yields $\hat{\xi} \approx 1.1$: close
to pure wealth-weighting ($\xi = 1$), consistent with the hypothesis
that visibility scales roughly linearly with wealth.

This calculation uses direct emigrants only.  Adding the 36 recent
heir-emigrants identified in \Cref{sec:heir_emigrants}---of which the
dated 2022 cases (H1--H3, H7)
carry approximately 73~bn~NOK---would raise the wealth-weighted
emigrant fraction substantially, but the correct accounting depends on
whether heir-emigrants are ``visible'' in the contagion sense.  A few
cases received extensive media coverage, while most moved quietly.  The
conservative approach is to treat $\hat{\xi} \approx 1.1$ as a lower
bound on the effective visibility weight, with the true value possibly
higher once intergenerational transfers are included.

\paragraph{The baseline emigration rate $\bar{\lambda}$.}
During the pre-reform period 2011--2017, no confirmed tax-motivated
departures occurred among the approximately 300 persons tracked
per year.  This places an upper bound on the pre-reform baseline rate:
$\bar{\lambda}\,(\tw^{\mathrm{pre}} - \tw^*) \lesssim
1/(7 \times 300) \approx 0.05\%$ per year---effectively zero.  During
the post-reform period (2021--2024), four events among approximately
420 persons give a flow rate of roughly 0.24\% per year.  The implied
jump in $\bar{\lambda}\,(\tw - \tw^*)$ is at least fivefold,
consistent with the reference-dependence channel~(F2) and the
composite tax-differential discussion of \Cref{sec:lambda}.

\paragraph{The Pareto tail exponent $\hat{\alpha}$.}
The Hill estimator at the 20\% tail fraction gives
$\hat{\alpha} \approx 1.3$, stable across all 15 years of the panel
(\Cref{sec:pareto_empirical}).  This parameter enters the
Fokker--Planck framework directly and is the best-identified quantity
in the model.

\paragraph{External benchmarks.}
The identified parameters can be cross-checked against comparable
estimates from other wealth-tax regimes.  The Swiss cantonal
decomposition of \citet{Brulhart2022} introduced in
\Cref{sec:intro}---which attributes roughly a quarter of the
behavioural response to migration---is the natural anchor, even
though the Swiss setting differs from Norway in rate level, federal
structure, and the absence of a comparable exit-tax regime.  The
contagion model nests smooth and tipping responses as limiting
cases: the Swiss experience operates in a regime of small,
persistent inter-cantonal differentials in which the contagion
channels (F2--F4) are largely inactive, and the resulting low
aggregate migration response corresponds to the low-emigration
equilibrium of the bistable model.  The Norwegian episode
corresponds to the opposite regime, in which the contagion channels
are jointly activated.  A single structural elasticity cannot span
both.

\begin{figure}[tp]
  \centering
  \includegraphics[width=\textwidth]{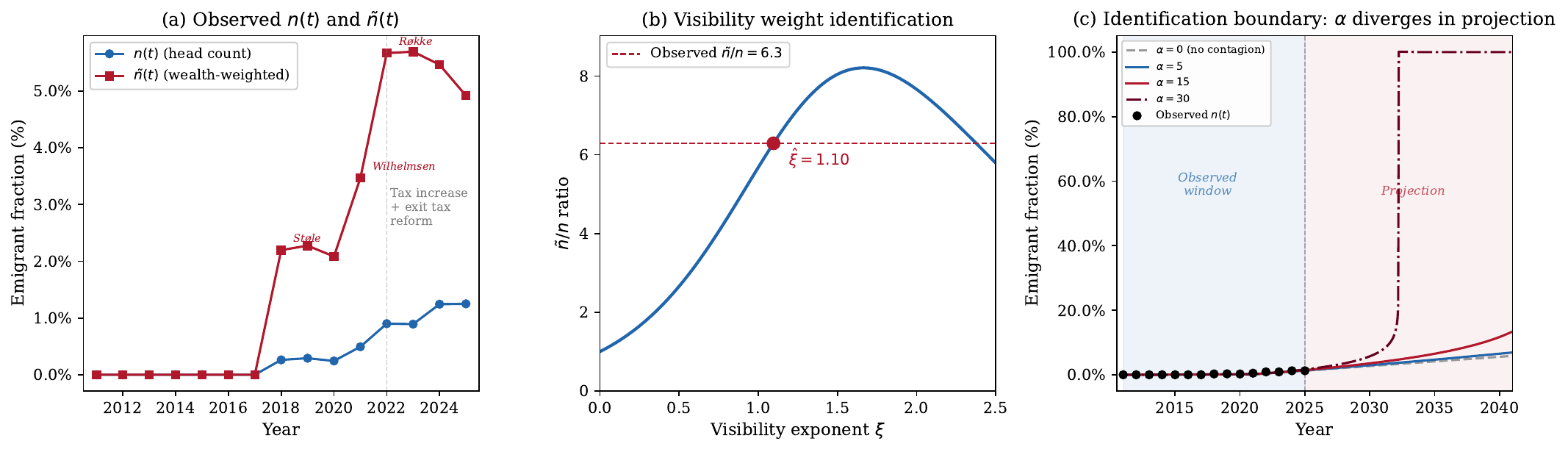}
  \caption{Data-calibrated parameters for the Kapital~400 panel
    (2011--2025), based on confirmed direct emigrants only
    (heir-emigrants excluded; see text for discussion).
    \textbf{(a)}~Observed emigrant count $n(t)$ and
    visibility-weighted emigrant fraction $\tilde{n}(t)$, showing the
    sharp post-reform acceleration driven by a small number of
    high-wealth departures.  \textbf{(b)}~Identification of the
    visibility weight: the ratio $\tilde{n}/n \approx 6.3$ pins down
    $\hat{\xi} \approx 1.1$ given the empirical Pareto tail.
    \textbf{(c)}~Identification boundary for the contagion
    strength~$\kappa$.  Within the observation window (shaded blue)
    all model trajectories are indistinguishable at $n < 1\%$,
    confirming that $\kappa$ is fundamentally unidentified from the
    available data.  The trajectories diverge only in projection
    (shaded red), illustrating the stakes of the unresolved
    uncertainty.}
  \label{fig:calibration}
\end{figure}

\subsection{What the data do not identify}\label{sec:unidentified}

The non-identification issues we face here are not peculiar to the
contagion model or to a single episode.  \citet{KlevenEtAl2020}, in
their canonical survey of the taxation-and-migration literature,
emphasise that migration elasticities are not structural primitives:
they depend on the tax differential, the anticipated persistence of
the reform, the stock of prior movers, the existence and bite of exit
taxes, and the broader policy regime.  Extrapolating a point estimate
from one reform or jurisdiction to another---as the DI exercise
does---therefore requires assumptions that the data cannot discipline.

Three parameters remain unidentified with the available data, and we
are explicit about why.

\paragraph{Contagion strength $\kappa$.}
Separating social contagion from a correlated response to a common
shock (the tax increase) requires either cross-sectional variation in
exposure to prior emigrants or cross-episode variation---neither of
which we have.  The observed $n(t)$ path is consistent both with
$\kappa = 0$ (no contagion; the five departures are independent
responses to the tax shock) and with $\kappa \gg 0$ (each departure
lowers the threshold for the next).  The S-shaped acceleration that
would distinguish contagion from a common shock is not resolvable at
annual frequency with five events.

\paragraph{Anticipation $\vartheta$.}
The exit-tax channel (F4) and the contagion channel (F1) both predict
acceleration during 2021--2022.  Since the wealth tax increase, the
exit-tax lapse abolition, and the first high-profile departures all
occurred in the same narrow window, the anticipation parameter~$\vartheta$
is confounded with~$\kappa$.  Identification would require an episode
where exit-tax rules changed \emph{without} a simultaneous wealth tax
change, or vice versa.

\paragraph{Regime hostility $\delta_h$.}
The 2021 government change and the 2022 tax increase are perfectly
collinear in our sample.  Separating hostility from the tax level
requires cross-government variation holding the tax rate constant---available in principle from the 2005 and 2013 Norwegian government
changes, but not in our Kapital~400 panel which begins in 2011.

\subsection{Identification boundaries and the DI extrapolation}

The non-identifiability of $\kappa$, $\vartheta$, and $\delta_h$ from a
single episode is not a deficiency of the contagion model---it is a
fundamental feature of the data-generating process.
\citeauthor{Blandhol2025}'s structural model faces the same
identification problem but resolves it by assumption: she imposes
$\kappa = 0$ (no contagion), $\vartheta = 0$ (no anticipation), and
$\delta_h = 0$ (no regime effect), reducing the model to independent
rational responses.  DI then extrapolates these point estimates
across countries.  Our contribution is to show that these restrictions
are not innocuous: the observed clustering, the timing relative to
exit-tax changes, and the $\tilde{n}/n$ ratio of~6.3 are all difficult to
reconcile with independent decision-making, even if the data cannot
pin down the exact parameter values.

The honest conclusion is a set of inequalities rather than point
estimates.  The emigration rate is bounded above by the observed
post-reform flow ($\sim$0.24\%/year among the ultra-wealthy); the
visibility weight is $\hat{\xi} \approx 1.1$; the Pareto exponent
is $\hat{\alpha} \approx 1.3$; and the contagion strength satisfies
$\kappa \geq 0$, with the data unable to rule out either extreme.
What the data \emph{do} rule out is the DI scaling exercise, which
requires all three non-identified parameters to be exactly zero and
the identified parameters to be portable across countries---a
conjunction of assumptions that is implausible given the evidence
assembled in \Cref{sec:data,sec:gap}.

\section{Implications for the Danish debate}\label{sec:policy}

The DI analysis \citep{DI2026} proceeds by taking Blandhol's
Norwegian estimates as given and scaling them to Danish conditions
through a simple ratio of wealth-tax-revenue-to-GDP.  The contagion
model exposes why this procedure is unreliable.

\begin{center}
\renewcommand{\arraystretch}{1.4}
\begin{tabular}{@{}p{5.5cm}p{5.5cm}@{}}
\toprule
\textbf{Standard model (DI assumption)} & \textbf{Contagion model} \\
\midrule
Independent emigration decisions &
  Emigration rate depends on visibility-weighted emigrant
  fraction $\tilde{n}$ \\
Responds to tax level $\tw$ &
  Responds to felt tax pressure:
  level $+$ change $+$ regime \\
Smooth elasticity &
  Tipping-point bifurcation \\
Linear scaling across countries &
  Non-scalable; depends on network structure, policy regime,
  anticipation \\
Large productivity haircut $\to$ GDP loss &
  Passive wealth moves; haircut $\approx 0$;
  GDP effect negligible \\
Time-invariant &
  Path-dependent; hysteresis; rush-for-the-door \\
Policy: adjust tax rate &
  Policy: manage the tipping point \\
\bottomrule
\end{tabular}
\end{center}

\medskip
Four concrete implications for the Danish proposal:
\begin{enumerate}[nosep]
  \item \textbf{Non-scalability.}  DI's linear extrapolation from
    Norway to Denmark is not justified.  The response is a nonlinear
    function of distance to the tipping point, which depends on
    country-specific network structure and social norms.  Denmark in
    2026 is not Norway in 2022: there is no recent tax increase to
    trigger reference-dependent pressure (F2), no regime shift
    comparable to the 2021 government change (F3), and no advanced
    exit-tax discussion creating anticipatory acceleration (F4).
  \item \textbf{Sequencing matters.}  Exit taxes enacted \emph{before}
    the cascade reduce $\bar{\lambda}$, raising
    $\kappa^{\mathrm{crit}}$ and preventing the bifurcation.  But
    \emph{announcing} exit taxes without enactment raises $p$,
    accelerating the very emigration it seeks to prevent.
  \item \textbf{Regime signaling.}  The policy hostility channel
    suggests that the \emph{framing} of tax policy matters as much as
    the rate.  A government that increases $\tw$ while otherwise
    signaling openness to capital may avoid triggering the cascade.
  \item \textbf{Hysteresis.}  Once the system tips, reducing $\tw$
    back to the pre-reform level may not restore the original
    equilibrium.  The return path requires $\ell$ to fall \emph{below}
    the forward tipping value---the system has memory.
\end{enumerate}

The DI report itself notes---as a reason to expect \emph{larger}
damage---that the proposed Danish wealth tax has a broader base than
the Norwegian one.  This observation, however, cuts both ways when
viewed through the Fokker--Planck lens.  To see why, it is useful to
compare the assessment regimes of the three relevant jurisdictions:
Norway, Denmark (as proposed), and Switzerland.

\paragraph{Norway.}
The Norwegian wealth tax base departs sharply from market value.
Unlisted shares are recorded at the company's fiscal book value of
equity, then multiplied by a statutory assessment fraction of 80\%
(\emph{verdsettingsrabatt} of 20\%).  Since book-to-market ratios for
unlisted firms are typically $\theta \approx 0.3$--$0.5$, the
effective assessment relative to true market wealth is
$\beta\,\theta \approx 0.24$--$0.40$---the tax base is roughly a
quarter to two-fifths of market value
(\Cref{thm:bv}).  Primary housing is assessed at only 25\% of
estimated market value up to NOK~10M (70\% above).  Debt, by
contrast, is deducted at face value.  The result is a strongly
anisotropic drift field: asset classes face different effective tax
rates, and leverage amplifies the wedge
(\Cref{eq:ch1}).  The drift-shift symmetry is
broken, creating incentives to tilt portfolios toward low-$\beta$
assets (unlisted equity, primary housing) and to lever up against
them.

\paragraph{Denmark (proposed).}
The Danish proposals are based on Danmarks Statistik's wealth
assessment (\emph{formueopg{\o}relse}), which records unlisted
businesses at an estimated market value derived from the relationship
between equity and share price for listed peers.  Listed shares enter
at market price; housing enters at the public property assessment
(\emph{ejendomsvurdering}); pension savings enter at their
accumulation value.  Under the Social Democrat proposal, the first
10M~DKK of primary-residence value (doubled for couples) is excluded
from the base.  The critical difference from Norway is that
business assets are \emph{not} discounted to book value.  If all
asset classes are assessed at or near market value, the assessment
fractions satisfy $\beta_i \approx 1$ for all~$i$, and the drift
shift (\Cref{eq:drift_shift}) becomes approximately uniform: the
tax is neutral in the Fokker--Planck sense.  A Danish wealth tax
would therefore avoid the portfolio distortions that plague the
Norwegian system---precisely the distortions that
\citet{Blandhol2025} observes in her event-study sample.

Paradoxically, the broader base that DI cites as a reason for
\emph{greater} damage is, in the symmetry framework, a reason for
\emph{less} distortion.  The Danish proposal eliminates Channel~1
symmetry breaking (\Cref{sec:drift_shift}): there is no built-in
incentive to shift wealth from listed to unlisted assets, from
financial to real estate assets, or to lever up against underassessed
collateral.  The economic response to such a tax should therefore
be governed by the smooth, neutral drift-shift---not by the
anisotropic portfolio reallocation that drives much of the Norwegian
experience.

\paragraph{Switzerland.}
The Swiss system occupies an intermediate position.  Listed shares
enter the cantonal wealth tax base at year-end market price.
Unlisted shares, however, are valued using the
\emph{Praktikermethode} prescribed by Federal Tax Administration
Circular~28: a weighted average of twice the capitalised earnings
value and once the substance value (book equity including latent
reserves),
\begin{equation}\label{eq:praktiker}
  V_{\mathrm{tax}} \;=\; \frac{2\,V_{\mathrm{earn}} + V_{\mathrm{sub}}}{3}\,.
\end{equation}
The earnings value is the average of recent net profits capitalised
at a reference rate (8.75\% for the 2024 tax year).  For profitable,
asset-light firms the resulting valuation can be substantially below
a market transaction price; for asset-heavy or loss-making firms it
converges toward book equity.  The assessment is thus neither pure
book value (Norway) nor pure market value (Danish proposal), but a
formulaic hybrid that introduces moderate---though not
extreme---anisotropy into the drift field.  Minority shareholders
(below 50\%) may additionally claim a 30\% discount.

The three regimes can be summarised in terms of the assessment
fraction~$\beta$ for unlisted business equity:
\begin{center}
\renewcommand{\arraystretch}{1.3}
\begin{tabular}{@{}lccc@{}}
\toprule
  & \textbf{Norway} & \textbf{Switzerland} & \textbf{Denmark (proposed)} \\
\midrule
Valuation basis & Book equity & Praktikermethode & Estimated market value \\
Statutory discount & 20\% & 0--30\% (minority) & 0\% \\
Effective $\beta$ (typical) & $\approx 0.3$ & $\approx 0.5$--$0.8$ & $\approx 1.0$ \\
Drift-shift symmetry & Strongly broken & Partially broken & Approximately preserved \\
\bottomrule
\end{tabular}
\end{center}

\noindent
The ordering is clear: the Norwegian system creates the largest
portfolio distortions and the strongest incentive to warehouse wealth
in underassessed vehicles; the Danish proposal creates the least.  It
is therefore not valid to extrapolate the Norwegian behavioural
response to Danish conditions, as the symmetry-breaking channel that
drives portfolio reallocation in Norway would be largely absent.

A further irony concerns rate levels.  The proposed Danish rate of
0.5\% is comparable to the \emph{effective} wealth tax levied in
Stadt Z\"urich.  Stadt Z\"urich applies a progressive cantonal
wealth tax reaching a simple rate of 3\textperthousand\ above
CHF~3.158M, which after the cantonal multiplier (95\% in 2026) and
the municipal multiplier (119\%) yields an effective marginal rate of
approximately 0.47\% at CHF~5M---nearly identical to the Danish
0.5\%.  While most Norwegian emigrants settle in low-tax cantons
(Zug, Schwyz, Lugano), some maintain offices or even residences in
Stadt Z\"urich, accepting a wealth tax comparable to the very rate
that dominates Danish political debate.  The critical difference lies
in the exemption threshold: Z\"urich grants only CHF~80\,000 per
individual (roughly 400\,000~DKK), whereas the Danish proposal
exempts 25M~DKK per single taxpayer.  A Dane with net wealth of
30M~DKK would pay 0.5\% on only 5M~DKK (25\,000~DKK per year),
while a Z\"urich resident with equivalent wealth (approximately
CHF~4M) pays the progressive levy on nearly the entire amount---a
burden several times larger.

The aggregate revenue comparison reinforces the point.  Norwegian
wealth tax revenue has grown to approximately NOK~34~billion in 2025,
or roughly 0.6\% of GDP.  Swiss cantonal and municipal wealth taxes
generate approximately 3.6\% of total tax revenue, corresponding to
about 1\% of GDP---more than any other OECD country.  The Danish
proposal, by contrast, targets 6--7~billion DKK from approximately
22\,000 taxpayers, or roughly 0.15\% of Danish GDP: a fraction of
what both Norway and Switzerland already collect.

\paragraph{The total tax burden, exit taxes, and the
\emph{fritaksmetoden}.}
The wealth tax alone does not explain the Norwegian emigration
incentive.  \citet{Blandhol2025} and the DI report focus exclusively
on the wealth tax rate, but the marginal Norwegian emigrant faces a
\emph{combined} burden of wealth tax, dividend tax, and latent
capital gains tax.  As \citet{BjerksundSchjelderup2021b} and
\citet{BjerksundHoplandSchjelderup2024} document, the
\emph{fritaksmetoden} (participation exemption) allows holding
companies to receive dividends and realise capital gains tax-free at
the corporate level, creating large pools of retained earnings with
deferred personal tax liabilities.  When the owner eventually
extracts these gains---whether as dividends (taxed at 37.84\% on
amounts above the risk-free return) or by selling shares---the
effective combined tax rate on corporate profits distributed to
individuals approaches 52\%.  Emigration offers a way to crystallise
these latent gains under a more favourable regime.

The exit tax is central to this calculus.  Until 29~November 2022,
Norwegian exit tax on unrealised share gains \emph{lapsed entirely}
if the emigrant waited five years without realising gains abroad.
This created an unambiguous crystallisation opportunity: emigrate,
wait, and the latent tax liability vanishes.  The timing is not
coincidental---the 2022 emigration wave peaked precisely as the
government announced it would close this window.  The anticipatory
acceleration channel (F4 in \Cref{sec:facts}) is visible in the
data: emigrants rushed to leave before the five-year lapse was
abolished.  The subsequent tightening on 20~March 2024 imposed a
twelve-year payment ceiling and triggered proportional exit-tax
payments on dividends received abroad, but the damage to the
credibility of the regime was done.

The legality of the tightened Norwegian rules is itself contested.
Free movement of persons is a fundamental pillar of the EEA
Agreement, and CJEU case law (notably \emph{National Grid Indus},
C-371/10) holds that exit taxes on unrealised gains are permissible
only if payment can be deferred without interest or security
requirements.  \citet{Banoun2025} has filed a formal complaint with
the EFTA Surveillance Authority arguing that Norway's post-2022 exit
tax rules---which require payment within twelve years regardless of
realisation, and trigger proportional payments on dividends received
abroad---violate Articles~28, 31, and 40 of the EEA Agreement.
The case (No.~93706) is under active examination by ESA
\citep{ESA2025}.  Switzerland, though not an EEA member, enjoys
equivalent free-movement protections through the bilateral Agreement
on the Free Movement of Persons (in force since 2002), which would
subject Norwegian exit tax rules to similar proportionality
constraints for moves to Switzerland.

Denmark has its own participation exemption---the EU
Parent-Subsidiary Directive makes this standard---and Danish holding
companies likewise accumulate untaxed retained earnings.  Denmark
also has an exit tax (\emph{fraflytterbeskatning}) that deems
unrealised gains realised upon emigration.  Crucially, however, for
moves within the EU and the Nordic area the Danish exit tax can be
deferred indefinitely without posting collateral and without interest
charges---precisely the design that CJEU jurisprudence requires.  The
Danish exit tax is thus consistent with free-movement principles but
is, for practical purposes, non-binding: a wealthy Dane can emigrate
to, say, Switzerland and defer the exit tax obligation indefinitely.

The implication cuts in two opposing directions.  On the one hand,
the absence of an effective exit tax means that a future Danish
wealth tax could trigger emigration without the friction that an
enforceable exit tax would provide.  On the other hand, the Norwegian
emigration wave was driven in large part by the \emph{interaction}
between the wealth tax increase, the accumulated
\emph{fritaksmetoden} deferral wedge, and a time-limited exit-tax
loophole that was about to close---a combination that has no analogue
in the Danish setting, where the proposed wealth tax would be new
and no comparable crystallisation window exists.  The DI report's
linear scaling from Norwegian to Danish conditions ignores all of
these asymmetries.

\paragraph{Coordination, embeddedness, and the transportability of
estimates.}
Three strands of the prior literature bear directly on the Danish
choice.  First, \citet{AgrawalEtAl2025}, exploiting the Madrid
regional wealth tax abolition, argue that unilateral wealth taxation
is structurally vulnerable to tax competition, and that the fiscal
externalities of top-wealth emigration extend well beyond the wealth
tax base itself.  Emigrating ultra-wealthy owners take with them
income, dividend, capital gains, and firm-level tax liabilities whose
combined size typically exceeds the direct wealth-tax loss.  The
implication is not that wealth taxation is futile but that its fiscal
effects are most naturally evaluated at the level of
\emph{coordinated} policy across jurisdictions that share a common
pool of mobile taxpayers.  Second, the embeddedness evidence of
\citet{Young2016} introduced in \Cref{sec:facts} requires an
additional qualification when invoked in cross-country comparisons:
the low US millionaire interstate migration rate reflects not only
local social capital and family attachment but also the distinctive
US citizenship-based tax regime.  Comparisons that invoke the US as
evidence of low mobility without acknowledging both factors will
systematically understate the migration margin in small, open,
domicile-based economies.  Third, the non-transportability of
migration elasticities established by \citet{KlevenEtAl2020} and
developed in \Cref{sec:unidentified} compounds the first two points:
estimates recovered from one reform or jurisdiction are not
structural objects that can be lifted into another regime.  Taken
together, these three observations reinforce the central claim of
the present paper.  The DI linear extrapolation from Norway to
Denmark rests on assumptions---no contagion, no anticipation, no
regime effect, portable elasticities, symmetric assessment---that
neither the Norwegian data nor the wider migration literature can
support.

The Danish proposal is, by international standards, a mild measure;
the political reaction it provokes reflects the contagion dynamics
this paper models rather than an objective assessment of its economic
burden.

\section{Conclusion}\label{sec:conclusion}

The Norwegian wealth-tax emigration episode has been treated in the
Danish policy debate as a natural experiment from which a portable
elasticity can be extracted and scaled to other countries.  This paper
has argued that the episode is better understood as a tipping event
driven by social contagion, reference-dependent tax pressure, policy
regime hostility, and exit-tax anticipation---four channels that
interact nonlinearly and produce path-dependent dynamics with
hysteresis.

The contagion model yields a sharp theoretical prediction: the
emigration system undergoes a saddle-node bifurcation at a critical
contagion strength $\kappa^{\mathrm{crit}} = 4$, above which two
stable equilibria coexist and the system can jump discontinuously
between them.  The Norwegian data are consistent with such a jump, but
the contagion strength itself is not identified from a single episode.
What the data do identify---the Pareto tail exponent
($\hat{\alpha} \approx 1.3$), the visibility weight
($\hat{\xi} \approx 1.1$), and the composition of emigrating
wealth---is sufficient to expose five independent failures in the
micro-to-macro extrapolation that yields the 1.3\% GDP estimate.  The
most consequential failure is representativeness: the wealth-weighted
integral that determines the GDP effect is dominated by individuals
who are either absent from the event-study sample or whose emigration
carried negligible productivity implications.  The 36 recent
heir-emigrants documented here---carrying 127~bn~NOK in pure economic
exposure with no attached human capital---are the clearest
illustration.

For Denmark, the implications are specific.  The proposed wealth tax
would operate on an approximately market-value base, preserving the
drift-shift symmetry that the Norwegian book-value system breaks.  It
would be introduced without a simultaneous regime shift, without a
closing exit-tax window, and without the accumulated deferral wedge
created by decades of the \emph{fritaksmetoden}.  None of the four
channels that pushed the Norwegian system past its tipping point would
be activated at comparable intensity.  The DI extrapolation, which
requires all four to be irrelevant, therefore answers the wrong
question: it estimates what would happen if Denmark replicated the
Norwegian episode, not what would happen if Denmark introduced a
wealth tax under Danish conditions.

The deeper lesson is methodological.  When a behavioural response
exhibits tipping-point dynamics, the standard toolkit of
micro-elasticities and linear scaling breaks down.  The relevant
policy question is not ``what is the elasticity?'' but ``how far is
the system from the tipping point?''---a question that depends on
country-specific network structure, institutional design, and policy
sequencing, and that cannot be answered by extrapolation from a single
foreign episode.

\subsection*{Acknowledgements}
The author acknowledges the use of Claude (Anthropic) for assistance with
literature review, \LaTeX{} typesetting, mathematical exposition, and
editorial refinement, and Lemma (Axiomatic AI) for review and proof
checking. All substantive arguments, economic reasoning, and conclusions
are the author's own.

\clearpage
\appendix

\section{Data sources and verification}\label{app:data}

This appendix documents the data sources, collection methodology, and
cross-checks that underlie the empirical analysis.

\subsection{The Kapital~400 panel}\label{app:panel}

The primary data source is the annual Kapital~400 list published by
Kapital magazine (Hegnar Media), which estimates the net worth
of the 400 wealthiest Norwegian-connected individuals.  The list data
are hosted on the Finansavisen web platform at
\texttt{finansavisen.no/kapital}.  We construct a
panel of 569 unique persons covering 2011--2025 (15 years,
approximately 5\,400 person-year observations) by scraping individual
person pages incrementally across multiple list years and merging with
the current (2025) edition.  The panel contains all 400 persons on the
current list plus 169 who appeared in earlier editions but have since
dropped off (due to death, wealth falling below the entry threshold,
or editorial reclassification).

For each individual-year observation the panel records:
\begin{enumerate}[nosep]
  \item a unique person identifier and URL slug (persistent across years),
  \item estimated net worth in NOK billions (the headline Kapital figure),
  \item taxable assets (\emph{ligningsformue}) where available from
    Skatteetaten filings, and
  \item tax paid, where reported.
\end{enumerate}
The net worth figures are Kapital's editorial estimates and should not be
confused with the tax authority's assessed wealth
(\emph{ligningsformue}), which follows statutory assessment rules
(book-value fractions, housing discounts, etc.).  The two series diverge
substantially for individuals with large unlisted equity holdings.

\paragraph{Coverage.}
For the 400 persons on the current (2025) list, coverage is complete
for 2020--2025 but incomplete for earlier years: the online
Kapital~Index retains historical person pages only for current list
members, so the backfilled data thins with age (168 entries for 2011,
rising steadily to 400 by 2020).  The 169 historical dropoffs
partially compensate: because their person pages were scraped while
they were still on the list, they contribute observations in years
where current-list backfill is sparse (e.g.\ 107 additional
observations in 2020).  Combined, the panel contains 410 observations
in 2020 and 444 in 2022, exceeding 400 when dropoffs overlap with
current members.  The summary statistics reported in the paper (total
wealth, Pareto tail exponents, concentration measures) are computed
from the full published list for each year, not from the panel
subset.  A separate summary file (\texttt{tail\_summary.csv}) records
the published aggregates.

\paragraph{Matching.}
Individuals are tracked across years using a combination of the
Kapital~Index person identifier and editorial slug.  Name changes
(marriage, deed poll) and family consolidations (where Kapital merges or
splits family entries) are handled manually.

\subsection{Emigrant identification}\label{app:emigrants}

We identify tax-motivated emigrants from the Kapital~400 panel using
four complementary methods:

\begin{enumerate}[nosep]
  \item \textbf{Ligningsformue drop (LIG).}  A sharp decline in taxable
    assets (typically to near zero or NaN) in consecutive publications
    signals that the individual is no longer filing Norwegian tax
    returns.  This is the strongest single indicator.
  \item \textbf{Tax paid decline (TAX).}  A progressive or sudden
    decline in reported tax paid, consistent with relocation to a
    lower-tax jurisdiction.
  \item \textbf{Residence flag (bosted).}  The Kapital~Index profile
    page reports the individual's stated country of residence.  We
    cross-check this against LIG and TAX signals.
  \item \textbf{Case study / news.}  For high-profile departures
    (R{\o}kke, Dahl, Moan), the emigration date and destination are
    confirmed from news reports.
\end{enumerate}

Each identified emigrant is classified as either \emph{direct} (the
individual emigrated personally) or \emph{heir} (wealth was transferred
to a next-generation family member who then emigrated or already resided
abroad).  The dating method is recorded for each entry.

\paragraph{Completeness.}
The emigrant list covers the universe of Kapital~400 members.  It does
not capture individuals who emigrated \emph{before} entering the list or
who never qualified for inclusion.  The Kapital~400 ``utflyttet''
(emigrated) flag identifies 67 persons in the 2025 list.  Of these, 23
are confirmed as recent tax-motivated direct emigrants in our data; the
remaining 44 are long-term expatriates (e.g.\ John Fredriksen, resident
in London/Cyprus since the 1990s; Torstein Hagen, resident in
Switzerland since the 2000s; Ole Andreas Halvorsen, resident in
Connecticut) whose emigration predates the 2022 policy episode and is
not tax-motivated in the sense modelled here.

Four of our identified direct emigrants no longer appear on the 2025
Kapital~400 list, having fallen below the entry threshold of
1.25~bn~NOK.  Their emigration status is confirmed from earlier list
years and news sources.

\subsection{Heir-emigrant identification}\label{app:heirs}

Heir-emigrants are identified from two sources:

\begin{enumerate}[nosep]
  \item A Finansavisen investigative article (August 2025) documenting
    intergenerational wealth transfers to emigrated heirs, listing 24
    named cases with estimated inherited wealth.
  \item The Kapital~300 list (October 2025), which for the first time
    included next-generation heirs as separate entries, yielding 44
    foreign-resident entries.
\end{enumerate}

The two sources overlap partially: 14 heirs appear in both, with some
name differences (the August list grouped siblings while the October list
split them into individual entries) and wealth updates (reflecting six
months of market movements and editorial revisions).  The paper's count
of 36 recent heir-emigrants is obtained by: (i)~splitting combined
entries into individuals, (ii)~merging the two sources, and
(iii)~excluding long-term foreign residents whose emigration predates
the policy episode and data artefacts (individuals flagged as
``utflyttet'' but resident in Norway).

\subsection{Cross-checks performed}\label{app:crosschecks}

A Python verification script (\texttt{data/verify\_data.py}) performs
the following automated cross-checks against published reference values:

\begin{enumerate}[nosep]
  \item \textbf{Aggregate totals.}  The panel total wealth for 2025
    matches the published figure of 2\,390.85~bn~NOK exactly.
  \item \textbf{Individual spot-checks.}  The 2025 wealth of 18 named
    individuals (including all top-10 and all identified emigrants still
    on the list) matches the interactive Kapital~Index to within 1\%.
  \item \textbf{Emigrant coverage.}  Of 27 identified direct emigrants,
    23 carry the ``utflyttet'' flag on the 2025 list; the remaining 4
    have fallen below the entry threshold.
  \item \textbf{Panel consistency.}  No entries with negative wealth;
    7 extreme year-on-year changes (${>}200\%$ growth or ${>}80\%$
    decline) are individually verified as genuine (e.g.\ inheritance
    events, IPOs, or editorial revaluations).
  \item \textbf{Pareto tail stability.}  The Hill estimator at the 20\%
    tail fraction yields a mean $\hat{\alpha} = 1.35$ with coefficient
    of variation 4.1\% across 2011--2025, confirming the stability
    claimed in \Cref{sec:pareto_empirical}.
  \item \textbf{Heir cross-check.}  The August and October heir lists
    are reconciled; wealth discrepancies exceeding 10\% (4 cases) are
    traced to entry splitting (combined sibling entries separated
    into individual records) or editorial revisions.
\end{enumerate}

The verification script is included in the replication archive.  The
underlying data files use the same coded identifiers as the paper
(\Cref{tab:emigrants,tab:heir_emigrants}); individual names are
replaced by codes so that the replication archive can be shared without
compromising the privacy protections described below.

\subsection{Data protection}\label{app:privacy}

The Kapital~400 list is published annually by Finansavisen as a
journalistic product; the wealth estimates, rankings, and emigration
flags used in this paper are drawn from that published source.
Processing of this publicly available data for academic research
purposes is permitted under Article~6(1)(e) of the General Data
Protection Regulation (task carried out in the public interest) and
the Norwegian \emph{personopplysningsloven} \S~8, which provides a
national legal basis for processing personal data for scientific
research purposes, subject to the safeguards in Article~89(1).

Consistent with the privacy protections in the body of the paper,
the appendix and replication archive observe the following principles:
\begin{enumerate}[nosep]
  \item Blandhol-window emigrants (2016--2020) and recent heir-emigrants
    are identified only by coded labels, not by name.
  \item Long-term expatriates and high-profile post-reform departures
    are named only where the individual's emigration is a matter of
    established public record (extensive news coverage, public
    statements, or listing on the published Kapital~400 ``utflyttet''
    register).
  \item The replication data files use the same coded identifiers as the
    published tables.  Researchers requiring access to the linked
    panel---which contains person identifiers matching the
    Kapital~Index---may request it from the authors subject to a
    data-use agreement specifying that individual-level records will
    not be republished.
\end{enumerate}

\end{document}